\documentstyle[pra,aps,amssymb,twocolumn,epsf]{revtex}

\def\H{{\cal H}}
\def\Cx{{\Bbb C}} 
\def\Rl{{\Bbb R}} 
\def\idty{{\openone}}
\def\tr{\mathop{\rm tr}\nolimits}

\def\V#1 {V_{(#1)}} 
\def\pr{{\bf P}}    
\def\wss{{\cal W}} 
\def\wssp{\wss^P}  

\def\wssc{\wss^{\rm cyc}}
\def\wssA{\wss^{(23)}}
\def\tsp{{\cal T}} 
\def\tspure{{\cal T}_{\rm pure}}
\def\bsp{{\cal B}} 
\def\ppt{{\cal P}}
\def\ket#1{\mid#1\rangle}
\def\ptr{^{T_1}}

\let\morE=\succ \def\dist{\Delta}

\def\ket #1{\vert #1\rangle}
\def\braket #1#2{\langle #1 \vert #2\rangle}
\def\ketbra #1#2{\vert #1\rangle \! \langle #2\vert}
\def\abs #1{\vert#1\vert}

\def\proof{\par\noindent{\it Proof:\ }}
\def\QED{\leavevmode\unskip\penalty9999 \hbox{}\nobreak\hfill
     \quad\hbox{\leavevmode  \hbox to.77778em{%
               \hfil\vrule   \vbox to.675em%
               {\hrule width.6em\vfil\hrule}\vrule\hfil}}
     \par\vskip24pt}
\def\norm #1{\Vert #1\Vert}
\def\ko #1#2{\left[#1,#2\right]_{-}}
\def\Null{{\bf 0}}

\def\pt{^{T_1}}

\newtheorem{The}{Theorem} 
\newtheorem{Lem}[The]{Lemma} 
\newtheorem{Cor}[The]{Corollary}

\title{Separability properties of tripartite states with
$\rm U\!\otimes\! U\!\otimes\! U$-symmetry}
 \author{T. Eggeling\thanks{Electronic Mail:
\tt{T.Eggeling@tu-bs.de}}{{}\quad and\ }
  R.~F. Werner\thanks{Electronic Mail: \tt{R.Werner@tu-bs.de}}
  \\[1ex]
  {\small Institut f{\"u}r Mathematische Physik, TU Braunschweig,}\\
  {\small Mendelssohnstr.3, 38106 Braunschweig, Germany.}}
\date{\today}

\begin{document}
\draft
\maketitle

\begin{abstract} We study separability properties in a
$5$-dimensional set of states of quantum systems composed of three
subsystems of equal but arbitrary finite Hilbert space dimension.
These are the states, which can be written as linear combinations
of permutation operators, or, equivalently, commute with unitaries
of the form $U\otimes U\otimes U$. We compute explicitly the
following subsets and their extreme points: (1) triseparable
states, which are convex combinations of triple tensor products,
(2) biseparable states, which are separable for a twofold
partition of the system, and (3) states with positive partial
transpose with respect to such a partition. Tripartite
entanglement is investigated in terms of the relative entropy of
tripartite entanglement and of the trace norm.
\end{abstract}

\pacs{03.65.Bz, 03.65.Ca, 89.70.+c}

\narrowtext

\section{Introduction} One of the difficulties in the theory of
entanglement is that state spaces are  usually fairly high
dimensional convex sets. Therefore, to explore in detail the
potential of entangled states one often has to rely on lower
dimensional ``laboratories''. An example of this was the role
played by a one-dimensional family of bipartite states \cite{W89},
which has come to be known as ``Werner states''. In this paper we
present a similar laboratory, designed for the study of
entanglement between three subsystems. The basic idea is rather
similar to \cite{W89}, and we believe this set shares many of the
virtues with its bipartite counterpart. Firstly, the states have
an explicit parameterization as linear combinations of permutation
operators. This is helpful for explicit computations. Secondly,
there is a ``twirl'' operation which brings an arbitrary
tripartite state to this special subset. This proved to be very
helpful for the discussion of entanglement distillation of
bipartite entanglement: the first useful distillation procedures
worked by starting with Werner states, applying a suitable
distillation operation, and then the twirl projection to come back
to the simple and well understood subset, thus allowing iteration
\cite{Pop1,Benn1}. Geometrically this means that the subset we
investigate is both a section of the state space by a plane and
the image of the state space under a projection. The basic
technique for getting such subsets is averaging over a symmetry
group of the entire state space. Such an averaging projection
preserves separability if it is an average only over local
(factorizing) unitaries. Of course, special subgroups might turn
out to be useful. For example, in a recent paper \cite{DCT99} a
class of tripartite ($n=3$) states was studied for dimension
$d=2,$ which is invariant under unitaries of the group of order 24
generated by $\sigma_1^{\otimes 3},$
$\sigma_3\otimes\idty\otimes\sigma_3,$
$\idty\otimes\sigma_3\otimes\sigma_3$ and
$\exp(i\pi\sigma_3/3)^{\otimes 3}.$

The third useful property of the states we study is that they can
be defined for systems of arbitrary finite Hilbert space dimension
$d$, leading to the same $5$-dimensional convex set for every $d$.
(This generalizes to an $(n!-1)$-dimensional set for $n$-partite
systems). Surprisingly, it turns out that the separability sets we
investigate are also independent of dimension.

We now describe the natural entanglement (or separability)
properties we will chart for these special states. Our
classification is similar to the one used in \cite{DCT99}, but
differs in that we do not artificially make the classes disjoint.

Of course, we can split the system into just two subsystems and
apply the usual separability/entanglement distinctions.  A split
$1\vert23$ then corresponds to the grouping of the Hilbert space
$\H_1\otimes\H_2\otimes\H_3$ into $\H_1\otimes(\H_2\otimes\H_3)$.
We call a density operator $\rho$ on this Hilbert space
{\it$1|23$-separable}, or just {\it biseparable}
if the partition is clear from the context, if we can write
\begin{equation}\label{def:bisep}
  \rho=\sum_\alpha \lambda_\alpha\ \rho^{(1)}_\alpha\otimes
          \rho^{(23)}_\alpha,
\end{equation}
with $\lambda_\alpha\geq0$ and density operators
$\rho^{(23)}_\alpha$ on $\H_2\otimes\H_3$. We will denote the
set of such $\rho$ by $\bsp_1$. This set will be
computed in Section~\ref{sec:bisep}. Furthermore, as it is a
necessary condition for biseparability (cf. Peres~\cite{Peres}),
we are going to look at those states $\rho$ having a {\it positive partial
transpose}  with regard to such a split, denoted by $\rho\in\ppt_1$.
Recall that the partial
transpose $A\mapsto A\ptr$ of operators on $\H_1\otimes\H_2$ is
defined by
\begin{equation}\label{eq:ppt}
  (\sum_\alpha A_\alpha\otimes B_\alpha)\ptr
      = \sum_\alpha A_\alpha^T\otimes B_\alpha,
\end{equation} where $A^T$ on the right hand side is the ordinary
transposition of matrices with respect to a fixed basis. It is
clear that $\bsp_1\subset\ppt_1$ holds, but as we will show in
Section~\ref{sec:ppt} by computing $\ppt_1$, this inclusion is
strict except for $d=2$.

As a genuinely ``tripartite'' notion of separability, we consider
states, called {\it triseparable} (or ``three-way classically
correlated''), which can be decomposed as
\begin{equation}\label{def:trisep}
  \rho=\sum_\alpha \lambda_\alpha\ \rho^{(1)}_\alpha\otimes
          \rho^{(2)}_\alpha\otimes\rho^{(3)}_\alpha,
\end{equation}
 where $\lambda_\alpha\geq0$, and the $\rho^{(i)}_\alpha$ are
density operators on the respective Hilbert spaces. The set of
such density operators will be denoted by $\tsp$.
Of course, we
may also consider states which are biseparable for all three
partitions. It is known \cite{BVMSST99} that this does not imply
triseparability, i.e.
$\tsp\subsetneqq(\bsp_1\cap\bsp_2\cap\bsp_3)$. Further examples
will be found below.

Since in this paper we will only be interested in a five
dimensional set $\wss$ of symmetric states (see the next section),
we will from now on use the symbols  $\tsp, \bsp_1$ and $\ppt_1$
only for the corresponding subsets of $\wss$.
\section{Definition and Main Results}\label{sec:def}
\subsection{$U\otimes U\otimes U$-invariant states: $\wss$}

Throughout we consider states on a Hilbert space of the form
$\H\otimes\H\otimes\H$, where $\H$ is a Hilbert space of finite
dimension $d$. The group of permutations on $3$ elements acts on
this space by unitary operators $V_\pi$, defined by %
$$ V_\pi\ \phi_1\otimes\phi_2\otimes\phi_3
   =\phi_{\pi^{-1}1}\otimes\phi_{\pi^{-1}2}\otimes\phi_{\pi^{-1}3}.
$$ For the six permutations $\pi$ we use cycle notation, so that
$\V12 $ is the permutation operator of the first two factors, and
$\V123 $ is the cyclic permutation taking $1$ to $2$. We denote by
``$dU$'' the normalized Haar measure on the unitary group of $\H$,
and define on the space of operators the operator %
\begin{equation}\label{def:pr}
  \pr \rho
    =\int\!dU\ (U\otimes U\otimes U)\rho\,(U\otimes U\otimes U)^*.
\end{equation}
Clearly, $\pr$ takes positive operators to positive operators (it
is even completely positive), and $\tr(\pr\rho)=\tr(\rho)$, i.e.,
$\pr$ maps density operators to density operators. We can now
define the set of states, which form the object of our
investigation.

\begin{Lem}\label{defWd}For an operator $\rho$ on
$\H\otimes\H\otimes\H$ the following conditions are equivalent:
\begin{enumerate}
 \item $(U\otimes U\otimes U)\rho=\rho(U\otimes U\otimes U)$ for all
unitary operators $U$ on $\H$.
 \item $\pr\rho=\rho$.
 \item $\rho=\sum_\pi\mu_\pi V_\pi$ with coefficients $\mu_\pi\in\Cx$.
\end{enumerate} The set of density operators satisfying these
conditions will be denoted by $\wss$.
 \end{Lem}

The equivalence of (1) and (2) is straightforward from the
invariance of the Haar measure. The implication
(3)$\Longrightarrow$(1) is trivial, because the permutation
operators clearly commute with operators of the form $(U\otimes
U\otimes U)$. The only non-trivial part is thus
(1)$\Longrightarrow$(3) which is, however, a standard result
(~\cite{Weyl} chap.IV) from representation theory. Of course, all
this works for any number of tensor factors.

The above Lemma does not address the question how to recognize
density matrices in terms of the six coefficients $\mu_\pi$.
Hermiticity requires $\mu_{\pi^{-1}}=\overline{\mu_\pi}$, leaving
effectively six real parameters. One more is fixed by
normalization, so that $\wss$ is embedded in a five dimensional
real vector space. In terms of the parameters $\mu_\pi$ positivity
is not easy to see. In order to get a better criterion it is best to study
the {\it algebra} of operators, which are linear combinations of
the permutations. The product of such operators can readily be
computed by using only the multiplication law for permutations.
The abstract algebra of formal linear combinations of group
elements (known as the group algebra) can be decomposed in terms
of the irreducible representations of the underlying group,
suggesting a basis which is much more handy for deciding
positivity. Again this step works for any number of factors, but
we carry it out only in the case $n=3$: We introduce the following
linear combinations of permutation operators:
\begin{mathletters}\label{V2R}
\begin{eqnarray}
  R_+\! &=& \frac{1}{6} \bigl(\idty+\V12 +\V23 +\V31 +\V123 +\V321 \bigr)\\
  R_-\! &=& \frac{1}{6} \bigl(\idty-\V12 -\V23 -\V31 +\V123 +\V321 \bigr)\\
  R_0 &=& \frac{1}{3} \Bigl(2\cdot\idty- \V123 -\V321 \Bigr)\\
  R_1 &=& \frac{1}{3} \Bigl(2\V23 -\V31 -\V12 \Bigr)\\
  R_2 &=& \frac{1}{\sqrt3}\ \Bigl(\V12 -\V31 \Bigr)\\
  R_3 &=& \frac{i}{\sqrt3}\ \Bigl(\V123 -\V321 \Bigr).
\end{eqnarray}
\end{mathletters}
Then $R_+,R_-,R_0$ are orthogonal projections
adding up to $\idty$, and commute with all permutations. This
means that they correspond to the irreducible representations of
the permutation group: $R_+$ and $R_-$ correspond to the two
one-dimensional representations (trivial and alternating
representation respectively), and these operators are indeed just the
orthogonal projections onto the symmetric and anti-symmetric
subspaces of $\H\otimes\H\otimes\H$ in the usual sense. Their
complement $R_0$ corresponds to a two dimensional representation,
which is hence isomorphic to the $2\times2$-matrices. The
operators $R_1,R_2,R_3$ act as the Pauli matrices of this
representation. In other words, the six hermitian operators
$R_+,R_-,R_0,R_1,R_2,R_3$ are characterized by the commutation
relations $R_iR_\pm=R_\pm R_i=0$, $R_i^2=R_0$, for $i=0,1,2,3$,
and $R_1R_2=iR_3$ with cyclic permutations.

Now every operator $\rho$ in the linear span of the permutations
can be decomposed into the orthogonal parts $R_+\rho$, $R_-\rho$,
and $R_0\rho$, and positivity of $\rho$ is equivalent to the
positivity of all three operators. This leads to the following
Lemma:

\begin{Lem}\label{le:range:ri}For any operator $\rho$ on
$\H\otimes\H\otimes\H$, define the six parameters
$r_k(\rho)=\tr(\rho R_k)$, for $k\in\lbrace+,-,0,1,2,3\rbrace$.
Then $r_k(\pr\rho)=r_k(\rho)$. Moreover, each $\rho\in\wss$ is
uniquely characterized by the tuple
$(r_+,r_-,r_0,r_1,r_2,r_3)\in\Rl^6$, and such a tuple belongs to a
density matrix $\rho\in\wss$ if and only if
\begin{eqnarray}
  r_+,r_-,r_0\geq0,&\qquad & r_++r_-+r_0=1 \nonumber\\
 {\rm and} &\qquad & r_1^2+r_2^2+r_3^2\leq r_0^2.\label{eq:range:ri}
\end{eqnarray}
\end{Lem}

Note that in this parameterization the set $\wss$ does not depend
on the dimension $d$ with one exception: for $d=2$ the
anti-symmetric projection $R_-$ is simply zero, so for qubits we
get the additional constraint $r_-=0$. If one considers a given
density operator $\rho$ as an operator $\rho'$ in
$\H'\otimes\H'\otimes\H'$ for a higher dimensional space
$\H'\supset\H$, by setting all ``new'' matrix elements equal to
zero, we will have $r_k(\rho)=r_k(\rho')$.

Taking $r_0=1-r_+-r_-$ to be redundant, we get a simple
representation of $\wss$ as a convex set in $5$ dimensions.
Unfortunately, $5$ dimensional sets are still not very amenable to
graphical representation. For visualizing the sets we are going to
describe analytically, we will therefore use suitable $2$ and $3$
dimensional representations. Again, we have the possibility of
using sections or projections of $\wss$, and we will emphasize
sections which can also be understood as projections.

\begin{figure}
\begin{center}
\epsfxsize=8.5cm \epsffile{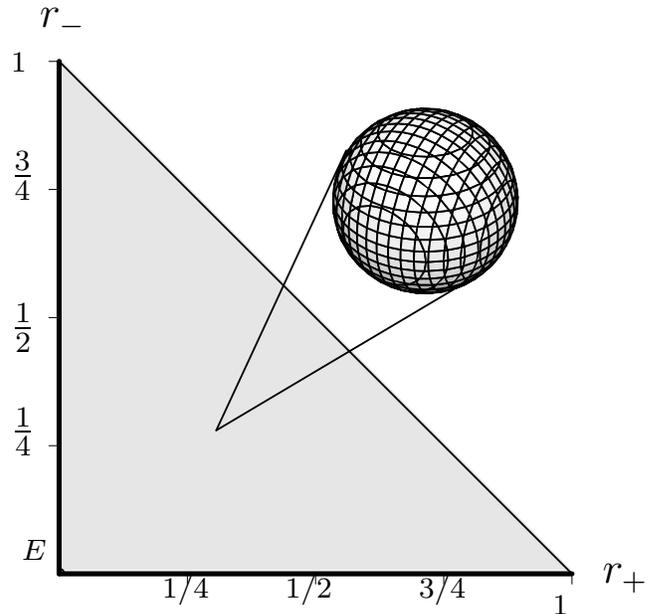} \caption{Description of
$\wss$ in terms of the triangle $\wss^P$ and the corresponding
Bloch sphere for each point in $\wss^P.$}\label{f1}
\end{center}
\end{figure}

The simplest example of this is to take the subset $\wssp\subset\wss$ of
states, which also commute with all permutations. The
corresponding projection is simply averaging with respect to
permutations. Clearly, $\wssp$ consists of those operators in
$\wss$, which are linear combinations of $R_+,R_-,R_0$ alone.
Taking $r_+$ and $r_-$ as coordinates we get the triangle in
Figure~\ref{f1}. Thus each point in this triangle represents a
density operator in $\wssp$. On the other hand, it represents the
set of states in $\wss$ projecting to it on permutation averaging:
this will be all states with the given values of $r_+$ and $r_-$
in the $6$-tuple, which therefore differ only in the values of
$r_1,r_2$, and $r_3$. Thus over every point of the triangle in
Figure~\ref{f1} we should imagine a Bloch sphere of radius $r_0$.

If more detail is required, we will also use three dimensional
sections and/or projections of a similar nature. For example, if
we average only over the permutations $\V23 $, we get the subset
$\wssA\subset\wss^{}$ with $r_2=r_3=0$ (see the dotted tetrahedron
in Figure~\ref{f9}). Averaging only over cyclic permutations, we
get the subset $\wssc\subset\wss^{}$ with $r_1=r_2=0$ (which gives
the same tetrahedron as $\wssA$ with $r_1$ substituted by $r_3.$).

We note for later use that the expectation values $r_k$ are {\it
not} the coefficients in the sum
 \begin{equation}\label{def:ci}
 \rho=\sum_{k=+,-,0,1,2,3}c_k R_k.
\end{equation}

These are related to the $r_k$ by the following dimension
dependent transformation (which is obtained by observing that
$\tr\idty=d^3$, $\tr(\V12 )=d^2$, and  $\tr(\V123 )=d$).

\begin{mathletters}
\label{§}
\begin{eqnarray}
  r_+\! &=& \frac{d}{6}(d^2+3d+2)\ c_+  \\
  r_-\! &=& \frac{d}{6}(d^2-3d+2)\ c_-  \\
  r_i &=& \frac{2d}{3}(d^2-1)\   c_i  \quad{\rm for}\
            i=0,1,2,3.
\end{eqnarray}
\end{mathletters}

\subsection{Overview of Main Results}\label{overview}

\begin{figure} \begin{center} \epsfxsize=8.5cm \epsffile{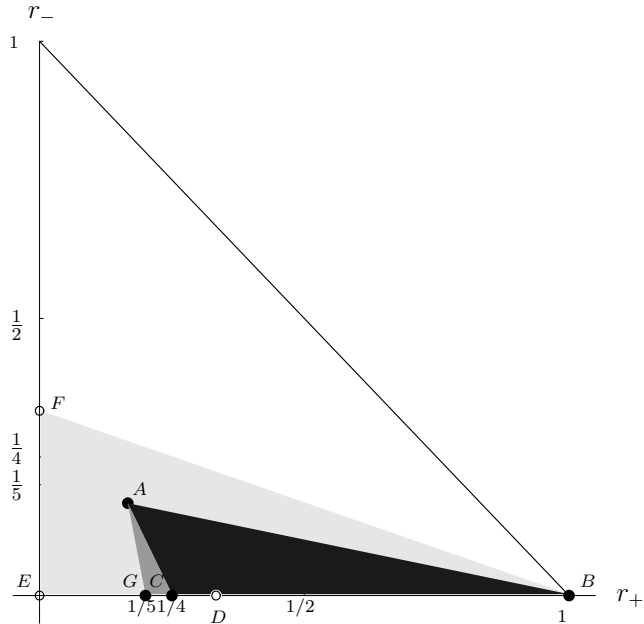}
\caption{Subsets of $\wssp$ with different separability
properties. Black: triseparable, dark grey: biseparable, light
grey: images of biseparable states under permutation averaging.
Special points explained in the text.}\label{f4} \end{center}
\end{figure}

An overview of the main results of this paper is given in
Figure~\ref{f4}. To keep the picture as simple as possible, we
have only depicted the set $\wssp$, i.e., the triangle in
Figure~\ref{f1}. Naturally, this reduction does not allow the
representation of our full results, i.e., the detailed structure
of the five-dimensional convex sets $\tsp,\bsp_1$, and $\ppt_1$,
which will be described in the corresponding sections. However, we
found this diagram quite useful as a basic map for not losing our
way in five dimensions, and hope it will similarly serve our
readers.

The shading in Figure~\ref{f4} marks different separability
properties, and the points labeled with capital letters arise by
projecting pure states with special properties with the twirl
projection (\ref{def:pr}). Some of these points (D,E,F) do not lie
in the plane $\wssp$, i.e, they have non-zero coordinates
$(r_1,r_2,r_3)$. They are represented by white circles, in
contrast to the black circles (A,B,C,G,H) representing permutation
invariant states in the plane $\wssp$.

The {\it triseparable} states correspond to the black triangle
$\triangle$(ABC). It is easy to see that any triseparable state
projected by permutation averaging to $\wssp$ is
again triseparable, i.e., the projection of $\tsp$ onto $\wssp$
coincides with $\tsp\cap\wssp$. The extreme points of this set are
\begin{eqnarray*}
 & A: & \ket{123} \longrightarrow (1/6,1/6,0,0,0) \\
 & B: & \ket{111} \longrightarrow (1,0,0,0,0) \\
 & C: & (\ket{111}-\sqrt{3}\ket{112}+\sqrt{3}\ket{121}-3\ket{122})/4 \\
 & & \qquad \qquad \qquad \qquad \qquad \longrightarrow (1/4,0,0,0,0)\;,
\end{eqnarray*}
 where the notation $\Psi\longrightarrow(r_+,r_-,r_1,r_2,r_3)$
indicates that the pure state $\ketbra\Psi\Psi$ is projected to
this point by $\pr$ from (\ref{def:pr}). In other words,
$\braket\Psi{R_k\Psi}=r_k$ for $k=+,-,1,2,3$. Note that all three
vectors given are product vectors, the one for C being the product
of three vectors in the ``Mercedes star'' configuration in the
plane, at angle $120^\circ$ from each other.

A quantitative description of the genuinely tripartite
entanglement of $\wss$ is given in section \ref{sec:relent} in
terms of the relative entropy and the trace norm.

The {\it biseparable} set $\bsp_1$ is not permutation invariant,
since the partition $1\vert23$ clearly is not. As a consequence,
the permutation average projecting $\wss$ onto $\wssp$ does not
map $\bsp_1$ into itself, and we have to distinguish in our
diagram between points $(r_+,r_-)$ such that $(r_+,r_-,0,0,0)$ is
biseparable (i.e., the {\it intersection} $\bsp_1\cap\wssp$), and
points $(r_+,r_-)$ such that for some suitable $(r_1,r_2,r_3)$ the
quintuple $(r_+,r_-,r_1,r_2,r_3)$ represents a point in $\bsp_1$,
(i.e., the {\it projection} of $\bsp_1$ onto $\wssp$). In
Figure~\ref{f4} the intersection is the triangle $\triangle$(GAB),
drawn in a darker shade of grey than the triangle
$\triangle$(EFB), which is the projection of the biseparable
subset $\bsp_1$.  Note that the shading reflects the inclusion
relations, i.e., triseparable states are, in particular,
biseparable, and the section of the biseparable set is contained
in its projection. Of course, the states in $\bsp_1\cap\wssp$ are
also biseparable for the other two partitions, since they are
permutation invariant. Similarly, the projections of $\bsp_2$ and
$\bsp_3$ onto $\wssp$ are the same.

Points of special interest for the biseparable set arise from the
following vectors:
 \begin{eqnarray*}
 & D: & \ket{122} \longrightarrow (1/3,0,2/3,0,0) \\
 & E: & (\ket{112}-\ket{121})/\sqrt{2} \longrightarrow (0,0,-1,0,0) \\
 & F: & (\ket{123}-\ket{132})/\sqrt{2} \longrightarrow (0,1/3,-2/3,0,0) \\
 & G: & (\ket{112}-\ket{121}-\sqrt{3}\ket{122})/\sqrt{5}
             \longrightarrow (1/5,0,0,0,0)\;.
\end{eqnarray*}
 Here the points B,D,E, and F are extreme points of $\bsp_1$, and
span a tetrahedron, which is equal to the subset $\bsp_1\cap\wssA$
of states invariant under the exchange $2\leftrightarrow3$. The
point G lies on the line connecting E and D, and is the unique
extreme point of $\bsp_1\cap\wssp$ which is not triseparable. In
this sense it represents an extreme case demonstrating the
inequality $\tsp\neq(\bsp_1\cap\bsp_2\cap\bsp_3)$.

The set $\ppt_1$ of states with {\it positive partial transpose}
with respect to the partition $1\vert23$ contains $\bsp_1$
strictly, but the difference cannot be seen in this diagram. In
fact, we will show in Section~\ref{sec:ppt} that even the
23-invariant subsets of $\ppt_1$ and $\bsp_1$ coincide, i.e.,
$\ppt_1\cap\wssA$ is spanned by the same four extreme points
B,D,E, and F.

As will be seen in Section~\ref{sec:ppt} there is a close
connection between the problems of finding $\ppt_1$ and finding
states invariant under averaging over all unitaries of the form
$\overline{U}\otimes U\otimes U$. It turns out that the sets of
triseparable and biseparable states commuting with such unitaries
can be obtained via a simple linear transformation from their
counterparts $\tsp\cap\wss$ and $\bsp_1\cap\wss$ computed in this
paper. This mapping and a sketch of the results is given in the
Appendix.
\section{Triseparable states: $\tsp$}\label{sec:trisep}

If $\rho$ is triseparable, hence has a decomposition of the
form~(\ref{def:trisep}), we may also find a decomposition in which
all factors $\rho^{(i)}_\alpha$ are pure, simply by decomposing
each of these density operators into pure ones. Applying to such a
decomposition the projection $\pr$ we find that
$\rho\in\tsp\subset\wss$ if and only if $\rho$ is a convex
combination of states of the form $\pr(\ketbra{\Psi}{\Psi}),$
where $\Psi=\psi_1\otimes\psi_2\otimes\psi_3$ is a normalized
product vector. Let us denote by $\tspure\subset\wss$ the set of
such states. Our strategy for determining $\tsp$ will be to first
get $\tspure$, and then to obtain $\tsp$ as its convex hull. The
resulting characterization of $\tsp$ is formulated in
Theorem~\ref{thm:tsp}.

Given a product vector $\Psi=\psi_1\otimes\psi_2\otimes\psi_3$, it
is easy to compute the projected state $\pr(\ketbra\Psi\Psi)$: By
Lemma~\ref{le:range:ri} one just has to compute the expectations of
the permutation operators. For example,
 $$\braket\Psi{\V12 \Psi}
   =\braket{\psi_1\otimes\psi_2\otimes\psi_3}
           {\psi_2\otimes\psi_1\otimes\psi_3}
   =\abs{\braket{\psi_1}{\psi_2}}^2.$$
In this way it is easily seen that the expectations of all
permutations are $\lbrace1,a_1,a_2,a_3,a_4+ia_5,a_4-ia_5\rbrace$,
where the $5$ real parameters are given by
\begin{mathletters}
\label{psi2a}
\begin{eqnarray}
   a_1 &=& \abs{\braket{\psi_2}{\psi_3}}^2 \\
   a_2 &=& \abs{\braket{\psi_3}{\psi_1}}^2 \\
   a_3 &=& \abs{\braket{\psi_1}{\psi_2}}^2 \\
   a_4 &=& \Re\! e\left(\braket{\psi_1}{\psi_2}
                \braket{\psi_2}{\psi_3}
                \braket{\psi_3}{\psi_1}\right) \\
   a_5 &=& \Im\! m\left(\braket{\psi_1}{\psi_2}
              \braket{\psi_2}{\psi_3}
              \braket{\psi_3}{\psi_1}\right).
\end{eqnarray}
\end{mathletters}
Since a pure state in $d$ dimensions (taken up to a factor) is
given by $2d-2$ real parameters, these $5$ quantities are a
considerable reduction from the $6(d-1)$ parameters determining
the three vectors $\psi_i$. However, they are still not
independent, due to the identity
\begin{equation}\label{ftrisep}
  f(a_1,a_2,a_3,a_4,a_5):=a_4^2+a_5^2-a_1a_2a_3=0.
\end{equation}
Since we want to determine $\tspure$ exactly, we also have to find
the exact range of these parameters, as the $\psi_i$ vary over all
unit vectors. This is done in the following Lemma.

\begin{Lem}\label{le:range:ai} A tuple
$(a_1,a_2,a_3,a_4,a_5)\in\Rl^5$ arises via equations~(\ref{psi2a})
from three unit vectors $\psi_1,\psi_2,\psi_3$ in a
$d$-dimensional Hilbert space ($d>3$), if and only if
equation~(\ref{ftrisep}) is satisfied,
 $0\leq a_i\leq1$ for $i=1,2,3$, and
\begin{equation}\label{eq:ai-range}
1-a_1-a_2-a_3 +2 a_4 \geq0.
\end{equation}
If $d=2$ the Lemma holds with last inequality replaced by
equality.
\end{Lem}

\proof
 Necessity of equation~(\ref{ftrisep}), and $0\leq a_i\leq1$ is
clear. Inequality~(\ref{eq:ai-range}) is just the condition that
the expectation of antisymmetric projection should be positive.
Since this projection vanishes for $d=2$, it is also clear that
equality must hold in this case.

Suppose now that $a_1,\ldots,a_5$ satisfying these constraints are
given. We have to reconstruct $\psi_1,\psi_2$, and $\psi_3$
satisfying equations~(\ref{psi2a}). These equations essentially
determine the $3\times3$-matrix $M_{ij}=\braket{\psi_i}{\psi_j}$
of scalar products. Of course, we already know the absolute values
of its entries (note $M_{ii}=1$). The phases are irrelevant up to
some extent: multiplying any row with a phase, and the
corresponding column with its complex conjugate will not change
the $a_i$ after equation~(\ref{psi2a}), and amounts to multiplying
one of the $\psi_i$ with a phase. Hence we may assume that the
scalar products $\braket{\psi_1}{\psi_2}$ and
$\braket{\psi_2}{\psi_3}$ are positive. The phase of the remaining
scalar product $\braket{\psi_3}{\psi_1}$ is then the same as the
phase of $a_4+ia_5$, hence $M$ is essentially uniquely determined
by the $a_i$.

Now a matrix $M$ is a matrix of scalar products if and only if it
is positive definite: on the one hand,
$\sum_{ij}\overline{u_i}u_jM_{ij}=\Vert\sum_iu_i\psi_i\Vert^2\geq0$.
On the other hand, we can construct a Hilbert space with such
scalar products as the space of formal linear combinations of
three vectors, with scalar products of basis vectors {\it defined}
by $M$. Positive definiteness of $M$ then ensures the positivity
of the norm in this new Hilbert space. The dimension of this space
is the rank of $M$ (number of linearly independent rows/columns).
So in the present case the dimension will be $3$ (but any larger
space will also contain appropriate vectors) or  $\leq2$, if $M$
is a singular matrix.

Positive definiteness of $M$ is equivalent to the positivity of
all subdeterminants. The diagonal elements are $1$, hence positive
anyway. Positivity of the three $2\times2$ subdeterminants is
equivalent to $a_i\leq1$ for $i=1,2,3$. Finally, the full
determinant of $M$, expressed in terms of the $a_i$ gives
expression (\ref{eq:ai-range}). It must be positive, and for
$d=2$ it must vanish, since $M$ is singular.\QED

Lemma~\ref{le:range:ai} describes the set $\tspure$ of
projected pure product states as a compact subset of the hypersurface
in $\Rl^5$ defined by equation~(\ref{ftrisep}). Computing the
convex hull of this set in $\Rl^5$ is the same as computing the
convex hull of $\tspure$, because the expectations of permutations
or the operators $R_k$ from (\ref{V2R}) are affine functions of
the $a_i$. Explicitly, the expectations
$r_k=\braket\Psi{R_k\Psi}$, $k=+,-,0,1,2,3$, which we have used as our
standard coordinates in $\wss$ are
\begin{eqnarray*}
 r_+\! &=& \frac{1}{6}\ (1+(a_1+a_2+a_3)+2a_4) \\
 r_-\! &=& \frac{1}{6}\ (1-(a_1+a_2+a_3)+2a_4) \\
 r_0 &=& \frac{2}{3}(1-a_4) \\
 r_1 &=& \frac{1}{3}\ (2a_1-a_2-a_3) \\
 r_2 &=& \frac{1}{\sqrt{3}}\ (a_3-a_2) \\
 r_3 &=& \frac{2}{\sqrt{3}}\ a_5.
\end{eqnarray*}

\begin{figure}
\begin{center}
\epsfxsize=8.5cm
\epsffile{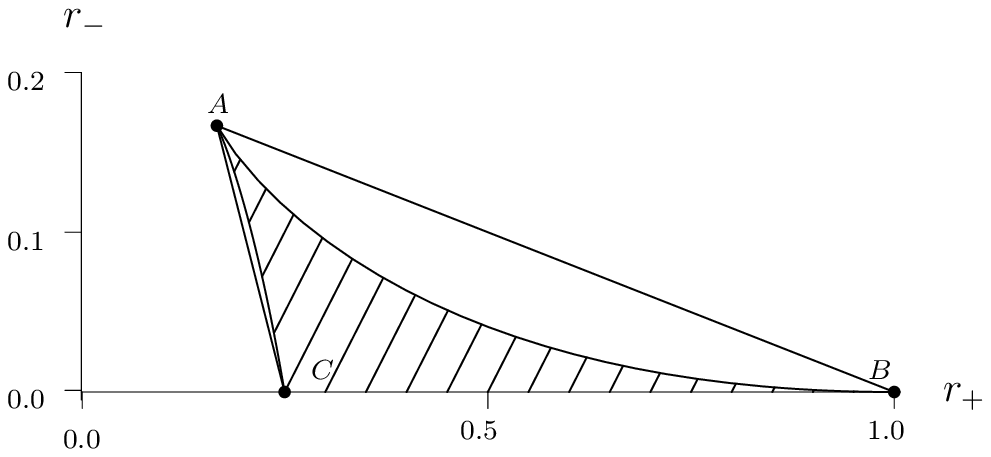}
\caption{Section of the set
$\tsp_{\rm pure}$ with $\wss^P$ and its convex hull.}\label{f5}
\end{center}
\end{figure}

We begin by computing the projection of $\tspure$ onto the
$(r_+,r_-)$-plane, by determining the possible range of the
combinations $m=(a_1+a_2+a_3)/3$ and $a_4$. By choosing phases for
the scalar products we can make $a_4$ vary in the range
$\abs{a_4}\leq (a_1a_2a_3)^{1/2}= g^{3/2}$, where $m$ and $g$ are
the arithmetic and the geometric mean of $a_1,a_2,a_3$. As is well
known, $g\leq m$, and equality holds if $a_1=a_2=a_3$. Hence the
projection of $\tspure$ is contained between the parameterized
lines
\begin{eqnarray*}
 r_+(m)&=&\frac16(1+3m \pm 2m^{2/3} )\\
 r_-(m)&=&\frac16(1-3m \pm 2m^{2/3} )\;.
\end{eqnarray*}
Plotting these curves gives
Figure~\ref{f5}. It is clear that the shape is not convex, and its
convex hull is the triangle $ABC$.

\begin{figure}
\begin{center}
\epsfxsize=8.5cm
\epsffile{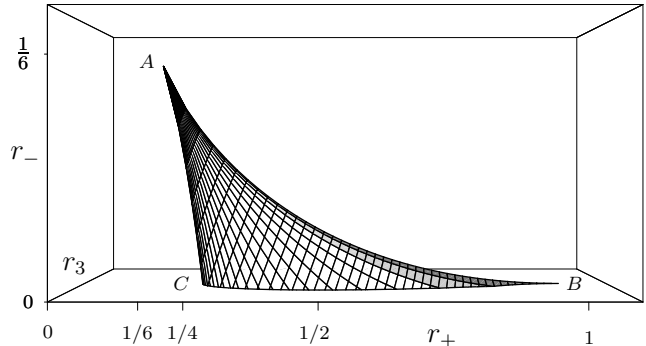}
\caption{Plot of the same section as above
making additional use of the coordinate $r_3.$}\label{f6a}
\end{center}
\end{figure}

A similar plot of the set $\tsp_{\rm pure}$ including one more
coordinate, $r_3$, is given in Figure~\ref{f6a}.

Again it is clear that no point on the surface can be an extreme
point of the convex hull of the surface, because the surface
``curves the wrong way''. This is the intuition behind the
following Lemma, by which we will show that also in the full
five-dimensional case the interior of $\tspure$ contains no
extreme points.

\begin{Lem}\label{lem:hyp} Let $N_f=\lbrace{x\in\Rl^n\mid f(x)=0\rbrace}$ be the
zero surface of a function $f\in{\cal C}^2(\Rl^n,\Rl)$, and
$K\subset\Rl^n$ a compact convex set. Let ${\cal U}$ be an open
ball around a point $x_h\in N_f$ such that $({\cal U}\cap
N_f)\subset K$, and suppose that $x_h$ is hyperbolic in the
following sense:
 $\nabla f(x_h)\neq0$, and the tangent plane through $x_h$
 contains two lines such that the second derivative of $f$ is
 strictly positive along one and strictly negative along the other.
\newline Then $x_h$ is not an extreme point of $K$. \end{Lem}

\proof Suppose $x_h$ is an extreme point of $K$. Then there must
be a supporting hyperplane, i.e., a hyperplane $H$ through $x_h$
such that $K$ lies entirely in one of the closed subspaces bounded
by $H$. We claim that this implies that $f$, restricted to $H$, has to
be either non-negative or non-positive in a
neighborhood of $x_h$.

Suppose to the contrary that there are points $x_+,x_-\in H\cap{\cal U}$
such that $f(x_+)>0>f(x_-)$. We may then connect $x_+$ and $x_-$ by a
continuous curve lying entirely in ${\cal U}$ and also in one of
the two open half spaces bounded by $H$. Since $f$ is continuous,
any such a curve must contain a point $y$ with $f(y)=0$, i.e.,
$y\in (N_f\cap{\cal U})\subset K$. Since we can choose either side
of $H$ for the connection, we find points $y\in K$ on both sides
of $H$, hence $H$ cannot be a supporting hyperplane.

This argument shows, in the first instance, that the only possible
supporting hyperplane at $x_h$ is the tangent hyperplane (look at
the Taylor approximation of $f$ to first order). Applying the
argument with the second order Taylor approximation, we find that
hyperbolic points cannot have supporting hyperplanes, hence cannot
be extremal. \QED

To apply this Lemma to the function $f$ from
equation~(\ref{ftrisep}), we have to pick two appropriate tangent
lines at any given point $\vec a=(a_1,a_2,a_3,a_4,a_5)$ on the
surface. We parameterize such lines as $\vec a+t\vec b$, $t\in\Rl$
so that
 $f(\vec a+t\vec b\,)=f(\vec a\,)+M t^2$. Two choices with opposite sign
of $M$ are
\begin{eqnarray*}
  \vec b\,&=& (0,0,0,a_5,-a_4),   \thinspace\qquad\qquad M=(a_4^2+a_5^2)\\
\text{ and } \vec b\,&=& (2a_1,2a_2,2a_3,3a_4,3a_5), \quad M=-3(a_4^2+a_5^2),
\end{eqnarray*}
where we have used the equation $f(\vec a\,)=0$ to evaluate the last
expression. Hence every point of the surface $N_f$ is hyperbolic.

By Lemma \ref{lem:hyp} we therefore only have to consider boundary
points of the surface, i.e., points for which at least one of the
inequalities in Lemma~\ref{le:range:ai} is equality.

Let us begin with the equalities $a_i=0,$ for at least one
$i\in\{1,2,3\}.$ Then we have $a_4=a_5=0$ by equation
(\ref{ftrisep}) and $0\leq a_j+a_k\leq 1$ ($j\neq k$) by equation
(\ref{eq:ai-range}). As we are looking for extremal points we are
left with the cases $a_j=a_k=0$ representing the triorthogonal
states\cite{footnote} (i.e. point
A=$(\frac{1}{6},\frac{1}{6},0,0,0)$ in the $r_i$'s) or
$a_j+a_k=1$. All such points satisfy $r_-=0$, hence they will be
in our general discussion of cases with $r_-=0$. The equalities
$a_i=1$ lead by (\ref{eq:ai-range}) to the inequality $0\leq 2
a_4-(a_j + a_k)$ and therefore to
\begin{eqnarray*}
a_4 & \geq & \frac{1}{2} (a_j+a_k)\geq\sqrt{a_ja_k}=\sqrt{a_ia_ja_k}\\
& = & \sqrt{a_4^2+a_5^2}\geq\sqrt{a_4^2}=\abs{a_4^{}}\geq a_4^{}.
\end{eqnarray*}
From this we can see $a_5=0,$ $a_j=a_k$ and $a_4=a_j=a_k.$ Once again this implies
$r_-=0$ so that this remains the only case to be checked.

For $r_-=0$, we can express the $a_i$ by $r_+,r_1,r_2,r_3$, and
solve equation~(\ref{ftrisep}) for $r_3$, obtaining a relation of
the form
\begin{equation}\label{r3trisep}
  r_3=\pm h(r_+,r_1,r_2)\;,
\end{equation}
where $h$ is the square root of a third order polynomial.
Equation~(\ref{r3trisep}) describes the surface of a convex set
iff $h$ is a concave function. This can be checked by verifying
that the Hessian of $h$ is everywhere negative semidefinite. Hence
all points in $\tspure$ with $r_-=0$ are extremal, and are
characterized by equation~(\ref{r3trisep}). This completes the
determination of extreme points of $\tsp$, summarized in the
following Theorem. It also contains the dual description of $\tsp$
in terms of inequalities.

\begin{figure}
\begin{center}
\epsfxsize=8.5cm
\epsffile{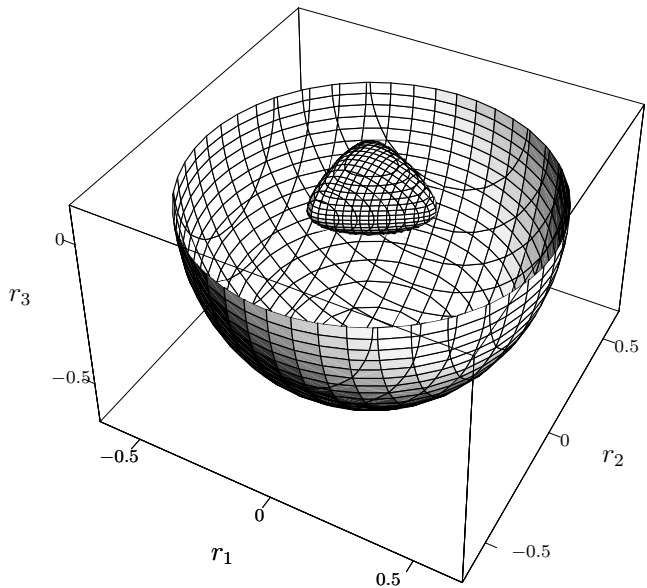}
\caption{Plotting the set $\tsp$ for the section $r_+=0.27,$ $r_-=0.1$ gives a
heart shaped surface with trigonal symmetry which is contained
 in the respective Bloch sphere.}\label{f6}
\end{center}
\end{figure}

\begin{The}\label{thm:tsp}The subset $\tsp\subset\wss$ of triseparable states
has the following extreme points, described here in terms of the
expectations $r_k=\tr(\rho R_k)$, $k=+,-,1,2,3\mathpunct:$
\begin{enumerate}
 \item $3r_3^2+(1-3 r_+)^2 = \\
 \hspace*{1cm} (r_1+r_+) \cdot (r_1-\sqrt{3}r_2-2 r_+)\cdot (r_1+\sqrt{3}r_2-2r_+)$\\
 and $r_-=0,$
 \item The point $A=(1/6,1/6,0,0,0)$.
\end{enumerate}
 A state $\rho\in\wss$ is triseparable if and only
if it corresponds to the point A or the following inequalities are
satisfied:
\begin{itemize}
\item[\rm (a)] $0\leq r_- < \frac{1}{6}$
\item[\rm (b)] $\frac{1}{4}(1-2r_-)\leq r_+ \leq 1-5r_-$
\item[\rm (c)] $(3r_3^2+[1-3 r_+ -3 r_-]^2)\cdot(1-6r_-)\leq \\
\hspace*{1cm} (r_1+r_+-r_-)\cdot\left((r_1-2 [r_+-r_-])^2 - 3r_2^2\right).$
\end{itemize}
\end{The}
These inequalities are obtained by projecting the given point onto
the hyperplane $r_-=0$ from point A, and to check whether the
projected point satisfies the inequality $\abs{r_3}\leq
h(r_+,r_1,r_2)$ with $h$ from equation~(\ref{r3trisep}). To get an
idea of the shape of $\tsp$ we compute the section with $r_+=0.27$
and $r_-=0.1$ (Figure~\ref{f6}).
\section{Relative Entropy of tripartite entanglement}\label{sec:relent}

Quantitative measures of bipartite entanglement and their
properties are a very active area of research at the moment. In
the tripartite case the difficulties in quantifying entanglement
begin already with the pure states, for which no canonical form as
simple as the Schmidt decomposition exists. One can, however,
extend the standard definition of the relation ``more entangled
than'' to tripartite states. It is clear what local quantum
operations should be in the multipartite case, and we can describe
classical communication between many partners in much the same way
as in the bipartite case. Once we fix the rules of classical
communication (e.g., ``each partner may broadcast her results to
all the others'') we will say that $\rho$ is more entangled than
$\sigma$, whenever we can reach $\sigma$ from $\rho$ by a sequence
of local operations and classical communication (LOCC), in which
case we will write $\rho\morE\sigma$.

A full characterization of this partial order relation is only
known in the case of bipartite pure states (Nielsen's Theorem
\cite{nielsen}). Even in the mixed bipartite case there is no
straightforward way of deciding whether one of two given density
operators is more entangled than the other. Hence we cannot hope
to give such a characterization in the tripartite case.
Nevertheless, the entanglement ordering is one of the features one
would like to explore and to chart in $\wss$. There are two ways
of approaching this: on the one hand, we may start from some state
$\rho\in\wss$, apply many LOCC operations to it, and see where we
end up. We can always assume the operation to end up in $\wss$,
because the twirl operation is itself a LOCC operation, which
involves the random choice of $U$ by any one of the partners, the
broadcasting of $U$ to the other two partners, and the unitary
transformation by $U$ at each of the sites. For an initial survey,
we may even study the relation in the permutation invariant
triangle $\wssp$ even though the permutation of sites is
definitely {\it not} a local operation. But if the inital state is
permutation invariant, and $T$ is any LOCC operation, involving
certain specified tasks for Alice, Bob and Charly, the three may
just throw dice to decide who is to take which role. With this
procedure they effectively get the permutation average of the
output state of $T$. With such studies, we get sufficient
conditions for $\rho\morE\sigma$.

In order to get necessary conditions the only approach is to find
functionals on the state space, which are monotone with respect to
entanglement ordering. Luckily, one of the ideas for getting such
monotones can be transferred from the bipartite case. Obviously,
the triseparable subset is invariant under LOCC operations, so the
distance to $\tsp$ is an entanglement monotone, provided the
distance functional has appropriate properties. One needs only one
condition for a function $\dist$ to define an appropriate
``distance'' $\dist(\rho,\sigma)$  between arbitrary states of the
same tripartite system:
 \begin{equation}\label{dmonotone}
  \dist(T\rho,T\sigma)\leq \dist(\rho,\sigma)
  \text{\ for any LOCC operation\ }T\;.
\end{equation}
 Then for the functional
 \begin{equation}\label{Ef}
E_\dist(\rho)=\inf\{\dist(\rho,\sigma)\vert \sigma\in{\tsp}\}
 \end{equation}
 we get the inequalities
 \begin{eqnarray}
 E_\dist(T\rho)&\leq&
  \inf\{\dist(T\rho,\sigma)\vert \sigma=T\sigma';\
                 \sigma'\in{\tsp}\} \\
  &=&\inf\{\dist(T\rho,T\sigma')\vert \sigma'\in{\tsp}\} \\
  &\leq&\inf\{\dist(\rho,\sigma')\vert \sigma'\in{\tsp}\}
  =E_\dist(\rho)\;.
 \end{eqnarray}
Hence $E_\dist$ is indeed a decreasing functional with respect to
the ordering $\morE$. Note that the only property of $\tsp$ needed
to show this is that it is mapped into itself under LOCC
operations. Any other set with that property (e.g., $\bsp_1$ or
$\ppt_1$) will also lead to an entanglement monotone.

Two natural choices for $\dist$ satisfy
requirement~(\ref{dmonotone}), and both of them satisfy it with
respect to arbirtrary operations $T$ (not just LOCC operations):
firstly the trace norm distance:
$\dist_1(\rho,\sigma)=\norm{\rho-\sigma}_1$, and the relative
entropy $\dist_S(\rho,\sigma)=S(\rho,\sigma)$, leading to
entanglement monotones we denote by $E_1$ and $E_S$, respectively.
In both cases, the actual computation of the distance for
$\rho,\sigma\in\wss$ is greatly simplified by the observation that
we may consider both $\rho$ and $\sigma$ as states (positive
normalized linear functionals) on the algebra generated by the
permutation operators, and that both the trace norm and the
relative entropy are naturally defined for such functionals
\cite{Petz}. Moreover, because the twirl (\ref{def:pr}) is a
conditional expectation the relative entropy of states in $\wss$
is independent of the algebra over which it is computed (cf. Thm.
1.13, \cite{Petz}). Now the $6$-dimensional algebra generated by
the permutations is independent of the dimension $d$, so that if
we parameterize $\rho$ and $\sigma$ by the expectations of $R_k$
as before, we find that the entanglement monotones $E_\dist$ are
independent of dimension. The expression for the relative entropy
involves, apart from the abelian summands the logarithm of a
$2\times2$-matrix, which can also be written explicitly in terms
of the parameters $r_k$ for the two states involved. The
variational problem~(\ref{Ef}) can be then solved numerically for
arbitrary states in $\wss$.

\begin{figure}
\begin{center}
\epsfxsize=8.5cm \epsffile{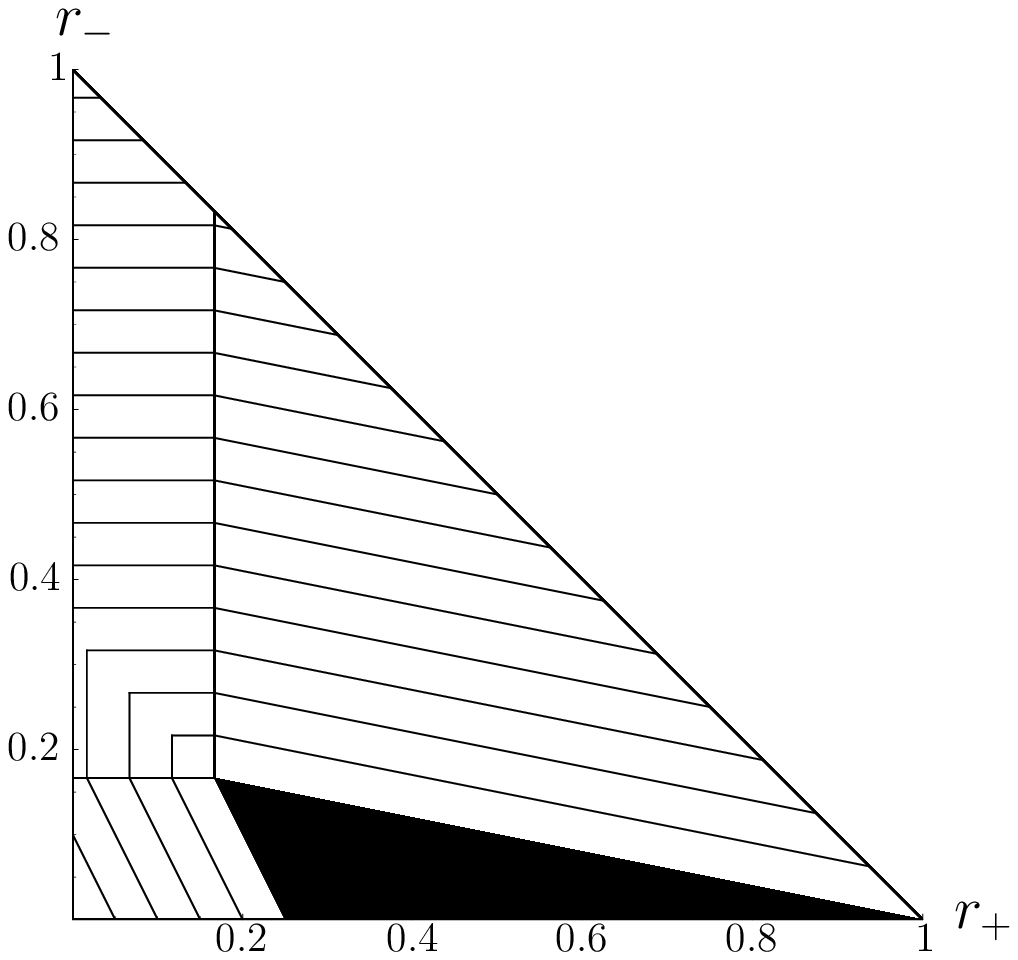}
\caption{Contour lines
over $\wssp$ for $E_1.$}\label{fhugo}
\end{center}
\end{figure}

\begin{figure}
\begin{center}
\epsfxsize=8.5cm \epsffile{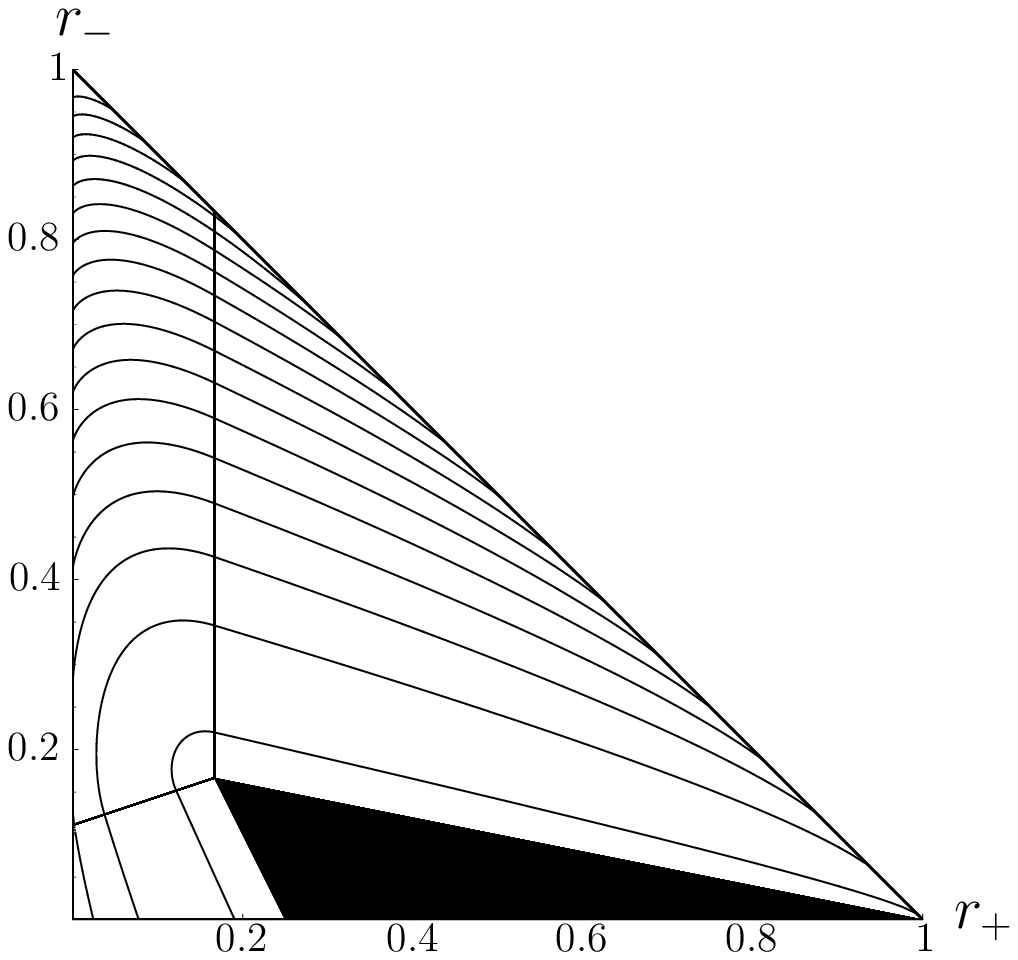} \caption{Contour lines
over $\wssp$ for $E_S.$}\label{fjoe}
\end{center}
\end{figure}

The contour lines over $\wssp$ of the resulting entanglement
monotones  are plotted in Figure~\ref{fhugo} for $E_1$, and  in
Figure~\ref{fjoe} for the relative entropy of tripartite
entanglement $E_S$. Note that the two neccessary conditions for
$\rho\morE\sigma$ expressed in these diagrams complement each
other. In order not to complicate these graphs we have not drawn
the simplest sufficient condition for entanglement ordering: from
any state $\rho$, any state lying on a straight line segment
ending in $\tsp$ is less entangled than $\rho$.

As a second section of interest we chose the plane
$r_-=0=r_1=r_2$, which is relevant for qubit systems.
Qualitatively, it gives the same picture of level lines wrapping
around the tripartite set.

\section{Biseparable states: $\bsp_1$}\label{sec:bisep}
In this section we are going to compute the set of biseparable
states with respect to the partition $1|23.$ The technique is
exactly the same as in the triseparable case: we first compute the
set $\bsp_{\rm pure}$ of states of the form
$\pr(\ketbra{\Psi}{\Psi})$ with $\ketbra{\Psi}{\Psi}$ biseparable,
i.e., $\Psi=\psi_1\otimes\psi_{2,3}$.  In a second step we get
$\bsp_1$ as the convex hull of $\bsp_{\rm pure}$.

We are free to apply to our vector $\Psi$ a $U\otimes U\otimes U$
rotation without changing the projection. In this way we may
choose $\psi_1=\ket1.$ Now the rotated state $\Psi'$ is of the
form $\Psi'=\sum_{i,j}\psi_{ij}\ket{1ij}.$ The expectations of
permutations of such a vector, like
$$\braket{\Psi'}{V_{(12)}\Psi'}
 =\sum_{i,j,k,l}\overline{\psi_{ij}}\psi_{kl}^{}\braket{1ij}{k1l}
 =\sum_{j}\abs{\psi_{1j}}^2$$
then depend linearly on the following real parameters:
\begin{mathletters}
\label{psi2c}
\begin{eqnarray}
c_0 &=& \abs{\psi_{11}}^2\\
c_1 &=& \sum_{j>1} \abs{\psi_{1j}}^2\\
c_2 &=& \sum_{i>1} \abs{\psi_{i1}}^2\\
c_3 &=& \sum_{i,j>1} \overline{\psi_{ij}}\psi_{ji}^{}\\
c_4+ic_5 &=& \sum_{j>1} \overline{\psi_{1j}}\psi_{j1}^{}\;.
\end{eqnarray}
\end{mathletters}
From this we obtain the following $r_k\mathpunct:$
\begin{eqnarray*}
r_+\! &=& \frac{1}{6}(1+5c_0+c_1+c_2+c_3+4c_4)\\
r_-\! &=& \frac{1}{6}(1-c_0-c_1-c_2-c_3)\\
r_0 &=& \frac{2}{3}(1-c_0-c_4)\\
r_1 &=& \frac{1}{3}(-c_1-c_2+2c_3+4c_4)\\
r_2 &=& \frac{c_1-c_2}{\sqrt{3}}\\
r_3 &=& \frac{2c_5}{\sqrt{3}}.
\end{eqnarray*}
As in the tripartite case we need to determine the exact range of
the parameters $c_i$. Let us assume $d>2$ for the moment. By the
definitions of $c_0,$ $c_1$ and $c_2$ we have
\begin{equation}\label{eq:c1}
c_0,c_1,c_2\geq 0.
\end{equation}
These parameters fix the weights of the blocks $(i=1,j=1)$,
$(i=1,j>1)$, and $(i>1,j=1)$ in the normalization sum
$\sum_{i,j=1}^d \abs{\psi_{ij}}^2=1$. $c_4+ic_5$ can be read as
the scalar product of two $(d-1)$-dimensional vectors
$\varphi_1=(\psi_{12},\dots,\psi_{1d})$ and
$\varphi_2=(\psi_{21},\dots,\psi_{d1})$ with norm squares
$\norm{\varphi_1}^2=c_1$ and $\norm{\varphi_2}^2=c_2.$ By the
Cauchy-Schwarz inequality we have:
\begin{equation}\label{eq:c2}
c_4^2+c_5^2=\abs{\braket{\varphi_1}{\varphi_2}}^2
     \leq \norm{\varphi_1}^2\norm{\varphi_2}^2=c_1c_2,
\end{equation}
and any value of $c_4+ic_5$ consistent with this can actually
occur.

We arrange the remaining $\psi_{ij}$ ($i,j>1$) into a
$(d-1)^2$-dimensional vector
$\widetilde\Psi=(\psi_{22},\dots,\psi_{2d},\psi_{32},\dots,\psi_{dd})$
with $\norm{\widetilde\Psi}^2=1-c_0-c_1-c_2$. On this
$(d-1)^2$-dimensional vector space, let $U$ denote the operator
swapping $\psi_{ij}$ and $\psi_{ji}$. Then
$c_3=\braket{\widetilde\Psi}{U\widetilde\Psi}$ is the expectation
of an hermitian operator with eigenvalues $\pm1$. Hence
\begin{eqnarray}\label{eq:c3}
\abs{c_3} \leq{\norm{\widetilde\Psi}^2}=1-c_0-c_1-c_2\;,
\end{eqnarray}
and all $c_3\in\Rl$ satisfying this inequality can occur.

Together with the obvious modifications in the case $d=2$, when
there is only one index $i>1$, we get the following Lemma:

\begin{Lem}\label{le:range:ci}
A tuple $(c_0,c_1,c_2,c_3,c_4,c_5)\in\Rl^6$ arises via
equations~(\ref{psi2c}) from a unit vector $\Psi$ in a
$d^2$-dimensional Hilbert space, if and only if equations
(\ref{eq:c1}), (\ref{eq:c2}) and (\ref{eq:c3}) are satisfied, and,
in the case $d=2,$ equality holds in (\ref{eq:c2}) and
(\ref{eq:c3}).
\end{Lem}

Let $\Gamma$ denote the set of tuples $(c_0,c_1,c_2,c_3,c_4,c_5)$
satisfying these constraints. The $r_k$ depend linearly on the
$c_i$, although the mapping is not one-to-one. Nevertheless any
extreme point of $\bsp_1$ must be the image of an extreme point of
the convex hull of $\Gamma$.

Hence we can proceed by first determine the extreme points of
$\Gamma$. Since the positive variables $c_0$, $\abs{c_3}$ and the
sum $(c_1+c_2)$ are only constrained by inequality~(\ref{eq:c3}),
every point in $\Gamma$ is a convex combination of tuples in which
only one of these is equal to $1$, and the other two vanish.
This gives the extreme points
\begin{enumerate}
\item $c_0=1 \Leftrightarrow \vec{r}=(1,0,0,0,0)\equiv B$
\item $c_3=+1\Leftrightarrow \vec{r}=(\frac{1}{3},0,\frac{2}{3},0,0)
       \equiv D$
\item $c_3=-1\Leftrightarrow \vec{r}=(0,\frac{1}{3},-\frac{2}{3},0,0)
        \equiv F$,
\end{enumerate}
and furthermore some points with $(c_1+c_2)=1$, $c_0=c_3=0$.
Eliminating $c_2=1-c_1$ we can write inequality~(\ref{eq:c2}) as
$c_4^2+c_5^2+(c_1-1/2)^2\leq1/4$. This is a ball with extreme
points parameterized by
\begin{eqnarray*}
c_0&=&0, \quad c_1=\frac{1+\cos (\vartheta)}{2},
       \qquad c_2=\frac{1-\cos (\vartheta)}{2}\\
c_3&=&0, \quad c_4=\frac{\sin (\vartheta)\cos (\varphi)}{2},
        \quad c_5=\frac{\sin (\vartheta)\sin (\varphi)}{2}
\end{eqnarray*}
with $\varphi,\vartheta\in [0,2\pi].$ By mapping this description
of $\Gamma$ to the $r_k$-parameterization we come to the following
Theorem:

\begin{The}\label{thm:bisep}The subset $\bsp_1\subset\wss$ of biseparable states with
respect to the partition $1|23$ has the following extreme points,
described here in terms of the expectations $r_k=\tr(\rho R_k)$,
$k=+,-,1,2,3$:
 \begin{enumerate}
 \item The sphere given by $\frac{1}{4}(3r_1+1)^2+3r_2^2+3r_3^2=1$ with $r_-=0$ and
 $r_+=(r_1+1)/{2}$ except for the point
 $(\frac{2}{3},0,\frac{1}{3},0,0)$, which is
    decomposable as $(\frac{2}{3},0,\frac{1}{3},0,0)=\frac{1}{2}(B+D)$
 \item The point $F=(0,\frac{1}{3},-\frac{2}{3},0,0)$
 \item The point $D=(\frac{1}{3},0,\frac{2}{3},0,0)$
 \item The point $B=(1,0,0,0,0).$
\end{enumerate}
A state $\rho\in\wss$ is biseparable with respect to the partition
$1|23$ if and only if it corresponds to the points F,B or D or the
following inequalities are satisfied:
\begin{itemize}
\item[\rm (a)] $0\leq r_- < \frac{1}{3}$
\item[\rm (b)] $-1<\frac{1+r_1-r_--2r_+}{1-3r_-}<1$
\item[\rm (c)] if $-1<\frac{1+r_1-r_--2r_+}{1-3r_-}\leq 0$ then
$$3r_2^2+3r_3^2+(1+2r_1^{}+r_-^{}-r_+^{})^2\leq (2+r_1^{}-4r_-^{}-2r_+^{})^2$$
\item[\rm (d)] if $0\leq\frac{1+r_1-r_--2r_+}{1-3r_-}<1$ then
$$3r_2^2+3r_3^2+(1-3r_-^{}-3r_+^{})^2\leq (r_1^{}+2r_-^{}-2r_+^{})^2.$$
\end{itemize}
\end{The}
We omit here again the computation of these inequalities from the known
extreme points. They can be obtained by projecting from the three
points $F,B$ and $D$ onto the sphere of extremal points.

\begin{figure}
\begin{center}
\epsfxsize=8.5cm
\epsffile{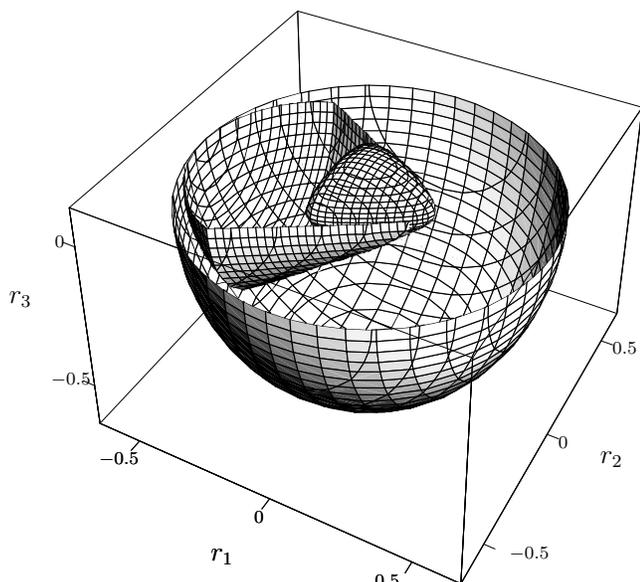}
\caption{Plot of the set $\bsp_1$ for $r_+=0.27$ and $r_-=0.1$ embedded
in the respective Bloch sphere together with $\tsp.$}\label{f8}
\end{center}
\end{figure}

The projection of the set $\bsp_1$ onto $\wss^P$ comes to be equal
to the projection of the set of pure $\bsp_1$-states and was
already shown in Figure~\ref{f4} together with the section
$\bsp_1\cap\wss^P.$ To compare $\bsp_1$ with $\tsp$ we plot again
the section with $r_+=0.27$ and $r_-=0.1$ (Figure~\ref{f8}).

To make the inclusion $\tsp\subsetneq(\bsp_1\cap\bsp_2\cap\bsp_3)$
we mentioned in the introduction more evident we can now compute
the sets $\bsp_2$ and $\bsp_3$ to build their intersection with
$\bsp_1.$ Due to the permutation symmetry of the three subsystems
we can rotate $\bsp_1$ by $\pm\frac{2\pi}{3}$ in the
$r_1$-$r_2$-plane instead. This leads to Figure~\ref{f8a}:

\begin{figure}
\begin{center}
\epsfxsize=8.5cm \epsffile{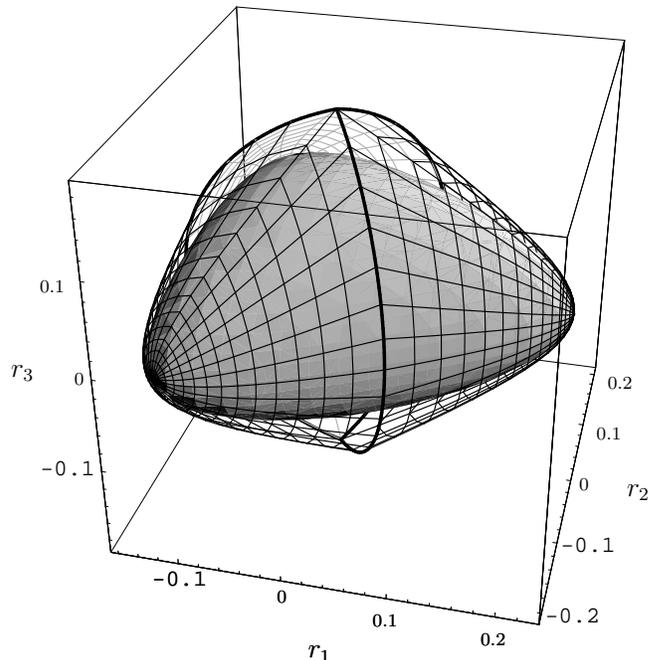} \caption{The intersection
$\bsp_1\cap\bsp_2\cap\bsp_3$ is shown as mesh on a transparent
surface allowing the set $\tsp$ to be seen. This plot is again
computed for the section $r_+=0.27$ and $r_-=0.1.$ The thick lines
indicate the intersection of two of the biseparable
sets.}\label{f8a}
\end{center}
\end{figure}
\section{Positive partial transposes: $\ppt_1$}\label{sec:ppt}

One of the interesting aspects in the theory of bipartite
entanglement to emerge in recent years was the consideration of
the partial transpose of the density matrix, and in particular the
positivity of the partial transpose. First, this positivity served
as a necessary condition for separability, which is even
sufficient in $2\otimes 2$ and  $2\otimes 3$ dimensions (the Peres
criterion \cite{Peres}). Moreover, it is a necessary condition for
undistillability, and here it comes much closer to sufficiency
even in general situations. Both aspects play a role in the
analysis of tripartite states. We will therefore describe in this
section the subset $\ppt_1\subset\wss$ of states with positive
$1$-transpose.

Since the dimensions for this bipartite system are $d\otimes d^2$,
positive partial transpose does not automatically imply
biseparability, i.e., the inclusion $\bsp_1\subset\ppt_1$ may be
strict. However, since we are considering a special class of
states it is also possible that in this class equality holds. This
does happen, for example, for the bipartite Werner states
\cite{Hor}. In the tripartite case we will see that
$\bsp_1=\ppt_1$ for $d=2$, but not for higher dimensions, although
the two sets come to be remarkably close (see Figure~\ref{f11}).
However, the exact description of $\ppt_1$ is also important for
distillation questions.

The partial transpose $A\mapsto A\pt$ of operators on
$\H_1\otimes\H_2$ was defined in Equation~(\ref{eq:ppt}). In a
tripartite system we take this operation to refer to the first of
the three tensor factors, and write $\rho\in\ppt_1$ if
$\rho\pt\geq0$.

\subsection{The algebra of partial transposes}

When $\rho$ is a linear combination of permutation operators as in
Lemma~\ref{defWd}, the partial transpose
$$ \rho\pt
   =\sum_{\pi} {\mu}_{\pi}V_{\pi}\pt$$
is likewise a linear combination of the six operators
$V_{\pi}\pt$, and we have to decide for which coefficients
${\mu}_{\pi}$ such an operator is positive. Since partial
transposition is {\it not} a homomorphism, it would appear that
the linear combinations of the $V_{\pi}\pt$ can be a fairly
arbitrary space of operators, and deciding positivity could be
quite difficult. However, it turns out that these linear
combinations do form an algebra, so after the introduction of the
right basis deciding positivity is just as easy as determining the
state space in Lemma~\ref{le:range:ri}.

The abstract reason for this ``happy coincidence'' is that the
operators $V_{\pi}\pt$ span the set of fixed points of an
averaging operation in much the same way as the permutations span
the set of fixed points of $\pr$. The corresponding averaging
operator is
\begin{equation}\label{def:prt}
  \widetilde{\pr} \rho
    =\int\!dU\ (\overline{U}\otimes U\otimes U)\rho\,(\overline{U}\otimes U\otimes U)^*.
\end{equation}
Its range consists of all operators commuting with all unitaries
of the form $\overline{U}\otimes U\otimes U$, hence is an algebra.
The following Lemma describes the relation between
$\widetilde{\pr}$ and $\pr$:

\begin{Lem}
Let $A$ be any hermitian operator, then
\begin{enumerate}\label{lem:prprt}
\item $\pr A=A\Leftrightarrow\widetilde{\pr}A\pt =A\pt$
\item $\left(\widetilde{\pr} A\right)\pt = \pr \left(A\pt\right).$
\end{enumerate}
\end{Lem}
\proof For any hermitian operator $A$ one has:
\begin{eqnarray*}
\widetilde{\pr} A=A &\Leftrightarrow &\ko{\overline{U}\otimes U\otimes U}{A}=\Null\\
&\Leftrightarrow & \ko{U\otimes U\otimes U}{A\pt}=\Null\\
&\Leftrightarrow &\pr A\pt=A\pt.
\end{eqnarray*}
Furthermore we can compute directly:
\begin{eqnarray*}
\pr A\pt & = & \int\!dU\ (U\otimes U\otimes U)A\pt\,(U\otimes U\otimes U)^*\\
& = & \int\!dU\ \bigl((\overline{U}\otimes U\otimes U)A\,(\overline{U}\otimes U\otimes U)^*\bigr)\pt\\
& = & \bigl(\widetilde{\pr}A\bigr)\pt.
\end{eqnarray*}
\QED

For deciding  positivity of partial transposes we need a concrete
form of the algebra spanned by the partial transposes of the
permutation operators. For example, we get
 \begin{eqnarray*}
   \V12 \pt&=&\sum_{ijk} \Bigl(\ketbra{ijk}{jik}\Bigr)\pt
          =\sum_{ijk}\ketbra{jjk}{iik}\\
         &=&(\ketbra\Phi\Phi)\otimes\idty,
\end{eqnarray*}
 where $\Phi=\sum_i\ket{ii}$ is a maximally entangled vector of
norm $d$. The partial transposes of the other permutations are
computed similarly. We can express all of them in terms of the
first two:
 \begin{equation}\label{pt-rules}
  X=\V12 \pt   \qquad {\rm and} \qquad V=\V23 \pt=\V23 %
\end{equation} as
$$ \idty\pt=\idty ,\quad \V13 \pt=VXV  ,\quad
  \V123 \pt=XV ,\quad  \V321 \pt=VX.
$$ Then these operators satisfy the relations $X^*=X,$ and $V^*=V$,
and
\begin{equation}\label{XVrel}
   X^2=dX \ ,\qquad  V^2=\idty \ ,\qquad XVX=X.
\end{equation} Due to these relations the set of linear combinations of the six
operators $\lbrace\idty,X,VXV,V,XV,VX\rbrace$ is closed under
adjoints and products. Positivity of such linear combinations, and
hence the positivity of all partial transposes of operators in
$\wss$ can therefore be decided by studying the abstract algebra
generated by two hermitian elements $X$ and $V$ satisfying
(\ref{XVrel}). As a six dimensional non-commutative C*-algebra it
is isomorphic to the algebra generated by the permutations, i.e.,
a sum of two one dimensional and a two dimensional matrix algebra.
But of course, the partial transpose operation mapping one into
the other is not a homomorphism.

From these considerations it is clear that all we have to do now
is to find a basis of the algebra generated by $X$ and $V$
analogous to the basis (\ref{V2R}). This sort of computation can
be quite painful, so we recommend the use of a symbolic algebra
package. The result is
\begin{mathletters}
\label{XV2S}
    \begin{eqnarray}
  S_+ &=&\frac{\idty+V}{2}\Bigl(\idty -\frac{2X}{d+1}\Bigr)
                    \frac{\idty+V}{2}\\%
  S_- &=&\frac{\idty-V}{2}\Bigl(\idty -\frac{2X}{d-1}\Bigr)
                    \frac{\idty-V}{2}\\%
  S_0 &=&\frac{1}{d^2-1}\
             \Bigl(d(X+VXV)-(XV+VX) \Bigr)\\%
  S_1 &=& \frac{1}{d^2-1}\ \Bigl(d(XV+VX)-(X+VXV) \Bigr)\\%
  S_2 &=& \frac{1}{\sqrt{d^2-1}}\ \Bigl(X-VXV \Bigr)\\%
  S_3 &=& \frac{i}{\sqrt{d^2-1}}\ \Bigl(XV-VX \Bigr).%
    \end{eqnarray}
\end{mathletters}
 These operators satisfy exactly the same relations as the $R_k$
from (\ref{V2R}) and we will denote the corresponding expectation
values by $s_k(\rho):=\tr(\rho S_k).$ The two projections $S_\pm$ correspond to the
two one-dimensional representations of the algebra, i.e., to the
two realizations of the relations by c-numbers, namely $X=0, V=1$
and $X=0,V=-1$.

\subsection{The $\V23 $-invariant case}
The simplest case is the $\V23 $-invariant subset of $\wss$
as it is a three dimensional object. In fact the $\V23 $-invariance
implies the conditions $\tr(\rho\V23 )\!=1,$ $\tr(\rho\V12 )\!=
\tr(\rho\V31 )$ and $\tr(\rho\V123 )\!=\tr(\rho\V321 ).$ Therefore we
have $r_2=r_3=0.$ In the same way we obtain for a $\V23 $-invariant state
$\rho\in\wss\pt$ the conditions $s_2=0$ and $s_3=0.$ Positivity of
a $\V23 $-invariant state in $\wss$ requires now $r_+\geq 0,$ $r_-\geq 0$
and $\abs{r_1}\leq r_0=1-r_+-r_-$ (cf. (\ref{eq:range:ri})) giving
raise to a tetrahedron bounded by the hyperplanes
\begin{eqnarray*}
(h_1)\ && r_+\!=0, \qquad (h_2)\ r_-=0, \qquad (h_3)\ r_1=
1-r_+-r_-\\ (h_4)\ && r_1= r_++r_--1
\end{eqnarray*}
and having the extreme points $P_1=(0,0,1),$ $P_2=(0,0,-1),$ $P_3=(0,1,0),$ and $P_4=(1,0,0).$
The same computation can be done on the partially transposed side
leading to the tetrahedron confined  by the hyperplanes
\begin{eqnarray*}
(h'_1)\ && s_+\!=0, \qquad (h'_2)\ s_-=0, \qquad (h'_3)\ s_1=1-s_+-s_-\\
(h'_4)\ && s_1= s_++s_--1.
\end{eqnarray*}
Using Lemma \ref{lem:prprt} we can express the $s_k$ by the $r_k$
of the corresponding $\wss$-state. Multiplying by positive
constants one gets an easier description of these hyperplanes:
\begin{eqnarray*}
(h'_1)\ && 2(1+r_1-r_--2r_+)+d(1+r_1-r_-+r_+)=0 \\ (h'_2)\ &&
2(-1+r_1+2r_-+r_+)+d(1-r_1+r_--r_+)=0 \\ (h'_3)\ &&
1-r_1-5r_--r_+=0\\ (h'_4)\ && 1+r_1-r_--5r_+=0.
\end{eqnarray*}
Its four extremal points are now
$Q_1=(\frac{2+d}{3},0,\frac{1-d}{3}),$
$Q_2=(0,\frac{2-d}{3},-\frac{1+d}{3}),$
$Q_3=(0,\frac{1}{3},-\frac{2}{3})$ and
$Q_4=(\frac{1}{3},0,\frac{2}{3}).$ Of course, these point have no
reason to correspond to positive states, and indeed only $Q_3$ and
$Q_4$ lie inside the state space, where $Q_1$ and $Q_4$ are
outside the state space for all $d$.

\begin{figure}
\begin{center}
\epsfxsize=8.5cm
\epsffile{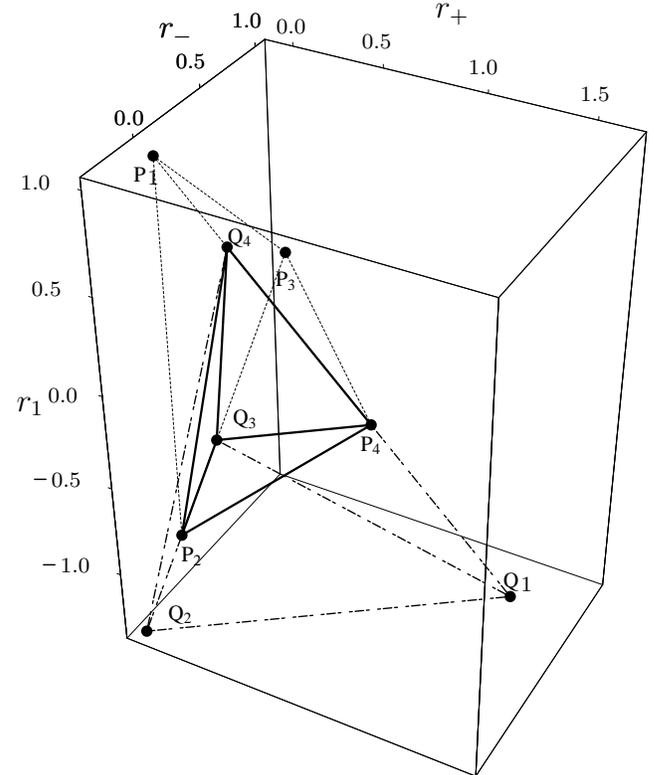}
\caption{The two positivity tetrahedra bounded by the $h_i$
(dotted) and the $h'_i$ (dashed) and the
intersection tetrahedron (solid lines) for $d=3.$}\label{f9}
\end{center}
\end{figure}

As we are looking for those $\V23 $-invariant $\wss$-states that
have positive partial transpose, i.e. that lie in $\ppt_A,$ we
have now to look at the intersection of these two tetrahedra. The
resulting object is again a tetrahedron as one can see in
Figure~\ref{f9}. This is due to the fact, that the extremal points
$P_i$ and $Q_i$ ($i=1,2,3,4$) lie on just two straight lines,
namely $\overline{P_1Q_4P_4Q_1}$ and $\overline{Q_2P_2Q_3P_3}.$
The intersection of the two tetrahedra is hence again a
tetrahedron, spanned by the extremal points $P_2$, $P_4$, $Q_3$
and $Q_4$ (called $E$, $B$, $F$, and $D$ in
Sections~\ref{overview} and \ref{sec:bisep}), and is thus
dimension independent. But it is easily verified from
Theorem~\ref{thm:bisep} that these four points are precisely the
extreme points of the $\V23 $-invariant part of $\bsp_1$. Since
$\bsp_1\subset\ppt_1$, we have shown the following:

\begin{Lem}\label{Lem:ppt23}
A $\V23 $-invariant $\wss$-state has a positive partial transpose
if and only if it is biseparable.
\end{Lem}

As we will see in the next subsection, the assumption of $\V23 $-invariance
is essential, i.e., the conclusion does not hold for general
$\wss$-states.

In order to see how $\V23 $-invariance helps, we conclude this
subsection with a direct proof of the above Lemma for $d=2$.
 If $\rho$ is a $\V23 $-invariant $\wss$-state, then we can
decompose it into the following sum
\begin{eqnarray*}
\rho\! &=& \!\frac{1}{4}\bigl(\idty+\V23
\bigr)\rho\bigl(\idty+\V23 \bigr)\!
   +\frac{1}{4}\bigl(\idty-\V23 \bigr)\rho\bigl(\idty-\V23 \bigr)\\
&=\mathpunct:& \rho^++\rho^-.
\end{eqnarray*}
It is now clear that $\rho$ has a positive partial transpose iff
both $\rho^+$ and $\rho^-$ each have a positive partial transpose.
$\rho^+$ denotes the $\V23 $-symmetric part of $\rho,$ $\rho^-$
the antisymmetric part. Thus we know that $\rho^+$ is a $2\times
3$ density operator and $\rho^-$ a $2\times 1.$ For these systems
the Peres criterion holds strictly \cite{Hor2}, i.e. states have a
positive partial transpose iff they are separable or in our case
biseparable over the $1\vert 23$ split. Biseparability of $\rho^+$
and $\rho^-$ is equivalent to the biseparability of $\rho$, which
proves the lemma.\QED

\subsection{The general case}
The positivity conditions for arbitrary linear combinations
of the operators $S_k$ give the following result:

\begin{Lem}\label{Lem:ppt}Let $\rho\in\wss$ be a density operator
with expectations $r_k=\tr(\rho R_k)$, $k=+,-,1,2,3$. Then the
partial transpose of $\rho$ with respect to the first tensor
factor is positive,i.e. $\rho\in\ppt_1,$ if and only if
\begin{mathletters}\label{thm:ppt}
\begin{eqnarray}
0&\leq&r_-\\
0&\leq&r_1-r_+-r_-+1\\
0&\leq&1-r_1-5r_--r_+\\
0&\leq&-1-r_1+r_-+5 r_+\\
r_2^2+r_3^2&\leq&R_1\label{thm:ppt:e}\\
r_2^2+r_3^2&\leq&R_2\label{thm:ppt:f}
\end{eqnarray}
\end{mathletters}
where
\begin{eqnarray*}
R_1&:=&(1-r_1-5 r_--r_+)(-1-r_1+r_-+5 r_+)/3\\
R_2&:=&(1-r_1-r_--r_+)(1+r_1-r_--r_+).
\end{eqnarray*}
\end{Lem}

\proof Recall that averaging with respect to $\V23 $ projects
$\ppt_1$ to the section of $\ppt_1$ with $r_2=r_3$. Therefore, the
inequalities describing the tetrahedron discussed in the last
subsection are optimal. These are the first four inequalities. We
therefore only have to describe the admissible set of $(r_2,r_3)$,
given $(r_+,r_-,r_1)$. There are two conditions to consider, one
from the positivity of $\rho$, and one from the positivity of
$\rho\pt$. As shown in the first subsection, both these
requirements have a very similar form, namely the positivity of an
element in an abstract algebra with two one-dimensional summands
and one summand isomorphic to the $2\times2$-matrices. Now in both
cases $(r_+,r_-,r_1)$ are readily seen to fix the weights of the
one-dimensional parts, as well as the trace and the expectation of
the first Pauli matrix for the $2\times2$-part. This leaves a
condition of the form $r_2^2+r_3^2\leq R$ in both cases. The two
conditions are given in the Lemma, where $R_2=(1-r_+-r_-)^2-r_1^2$
expresses the requirement $\rho\geq0$. The condition
(\ref{thm:ppt:e}) is obtained from $\rho\pt\geq0$ by expressing
$\rho\pt$ in the basis $S_k$, and applying the same criterion to
the expectations $s_k$.\QED

\begin{figure}
\begin{center}
\epsfxsize=8.5cm \epsffile{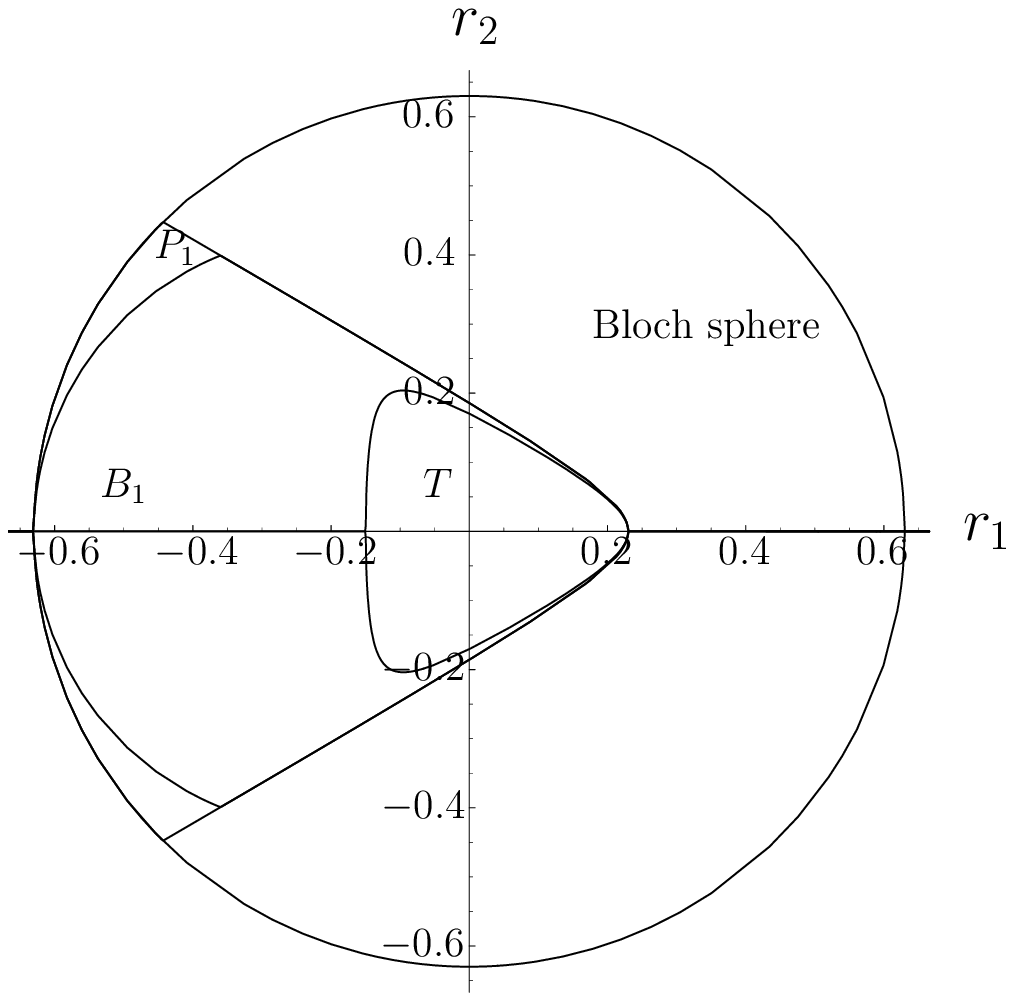} \caption{Plot of the Bloch
sphere, $\tsp,$ $\bsp_1$ and $\ppt_1$ for $r_+=0.27,$ $r_-=0.1$
and $r_3=0.$}\label{f7}
\end{center}
\end{figure}

According to this Lemma the set $\ppt_1$ can be visualized as
follows: firstly, one has to fix a point $(r_+,r_-)$ in the
permutation invariant triangle (Figure~\ref{f4}). The possible
choices of $(r_1,r_2,r_3)$ can then be seen from Figure~\ref{f8}.
Apart from the heart shaped tripartite set in the center this
Figure contains three quadratic surfaces: the Bloch sphere, and
the two surfaces bounding $\bsp_1$. Comparing condition (d) of
Theorem~\ref{thm:bisep} and the expression for $R_1$ given in the
above Lemma, we find that both constraints are given by the same
hyperboloid, the one wrapped around the tripartite set in
Figure~\ref{f8}. Hence in that Figure we can readily find $\ppt_1$
by extending this hyperboloid all the way to the Bloch sphere, and
taking the intersection. This is shown in Figure~\ref{f7}, in the
section $r_3=0$.

Figure~\ref{f7} shows the generic situation with $r_-\neq0$. When
$r_-=0$, in particular for systems of three qubits, the boundary
ellipsoid of $\bsp_1$, described by condition (c) of
Theorem~\ref{thm:bisep}, coalesces with the Bloch sphere. This
leads to another instance where the Peres-Horodecki criterion for
separability holds:

\begin{Cor}\label{pptQbit}
The intersections of $\bsp_1$ and $\ppt_1$ with the plane $r_-=0$
coincide. In particular, for $3$-qubit $\wss$-states,
biseparability is equivalent to the positivity of the partial
transpose.
\end{Cor}

We conclude this section by the explicit determination of the
extreme points of $\ppt_1$. From Figure~\ref{f7} it might appear
that all points on the quadratic surfaces bounding $\ppt_1$ might
be extremal. But this is misleading, because we also have to take
into account the possibility of decompositions with different
values of $(r_+,r_-)$. In fact, for the inequalities arising from
$\rho\geq0$ it is evident that generically such decompositions are possible:
given any $(r_+,r_-,r_1,r_2,r_3)$, which lies on the Bloch sphere in
Figure~\ref{f7}, we can just change the
weights of the three blocks in the block decomposition of
$\rho$ according to
$(\lambda_+r_+,\lambda_-r_-,\lambda_0r_1,\lambda_0r_2,\lambda_0r_3))$,
as long as the $\lambda_\alpha$ are positive, and the normalization
$\lambda_+r_++\lambda_-r_-+\lambda_0(1-r_+-r_-)=1$ is respected.
This leaves a two dimensional affine manifold through
$(r_+,r_-,r_1,r_2,r_3)$. Hence, unless other conditions
constraining $\ppt_1$ prevent the indicated decompositions no such
point will be extremal. Of course, the second constraint
(\ref{thm:ppt:e}) has the same structure, because the algebra of
partial transposes is isomorphic to the algebra generated by the
states. Hence in Figure~\ref{f7} only the points in the
intersection of the hyperboloid and the Bloch sphere remain as
candidates for extreme points. This is analogous to the
extreme points of $\bsp_1$, which also consist of the intersection
of two quadratic surfaces in Figure~\ref{f7}. For $\ppt_1$ we get

\begin{The}\label{thm:pptex}
The subset $\ppt_1\subset\wss$ of $\wss$-states with positive
1-transpose has the following extreme points, described here in
terms of the expectations $r_k=\tr(\rho R_k)$, $k=+,-,1,2,3$:
\begin{enumerate}
\item The points $P_2,$ $Q_3,$ $P_4,$ and $Q_4$, which also span
the $\V23 $-invariant part of $\ppt_1$.
\item the remaining extreme points of $\bsp_1$, which form a
sphere in the $r_-=0$ plane (cf. Theorem~\ref{thm:bisep}).
\item The points for which $(r_+,r_-,r_1,0,0)$ lie in the interior
of the $\V23 $-invariant tetrahedron, and for which inequalities
(\ref{thm:ppt:e}) and (\ref{thm:ppt:f}) are both satisfied with
equality.
\end{enumerate}
\end{The}

\proof Let us first discuss the periphery of the tetrahedron.
Every face of the tetrahedron corresponds to a face of $\ppt_1$,
namely the face of points projecting to it upon $\V23 $-averaging.
In Lemma~\ref{Lem:ppt} this corresponds to the subsets for which
one of the linear inequalities (\ref{thm:ppt}a) to
(\ref{thm:ppt}d) is equality.
We will show first that each of these faces is actually contained
in $\bsp_1$. Indeed, when (\ref{thm:ppt}b), (\ref{thm:ppt}c) or
(\ref{thm:ppt}d) are equalities, one of the factors in $R_1$ or
$R_2$ vanishes, forcing $r_2=r_3=0$, reducing our claim to
Lemma~\ref{Lem:ppt23}. When (\ref{thm:ppt}a) is equality, i.e.,
$r_-=0$, the claim is contained in Corollary~\ref{pptQbit}.

Now a point of $\ppt_1$ contained in one of these faces can
only have decompositions in the same face, hence in $\bsp_1$,
hence for such a point extremality in $\ppt_1$ and extremality in
$\bsp_1$ are equivalent.

It remains to show item 3 of the Theorem, i.e., to characterize
the extreme points of $\ppt_1$, whose $\V23 $-averages fall in the
interior of the tetrahedron. From the arguments preceding the
Theorem it is clear that points for which only one of the
inequalities (\ref{thm:ppt:e}) and (\ref{thm:ppt:f}) are
equalities cannot be extremal, since the surfaces defined by these
equations contain straight lines. Therefore, the condition stated
in the Theorem is necessary for a point to be extremal. It remains
to show that none of the points with $R_1=R_2$ can be decomposed
in a proper convex combination.

Let us denote by $M_1$ (resp. $M_2$) the set of those points in
the interior of the tetrahedron such that $R_1\leq R_2$ (resp. $R_2\leq
R_1$). The intersection $M_*=M_1\cap M_2$ of these sets is described by
the condition $R_1=R_2$, or explicitly
\begin{equation}\label{cut-edge}
 r_1^2 + 3r_- + r_1 r_- - 2r_-^2 + 3r_+ - r_1r_+ - 8r_- r_+ -2r_+^2
 \!=\!1.
\end{equation}

\begin{figure}
\begin{center}
\epsfxsize=8.5cm \epsffile{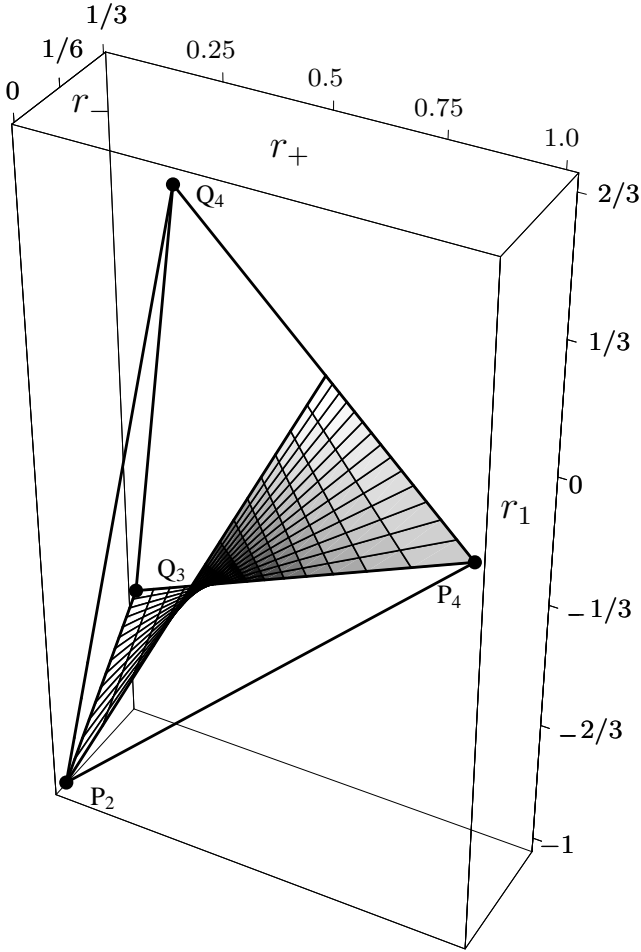} \caption{Section of the
intersecting tetrahedron with the separating one-leaf
hyperboloid.}\label{f10}
\end{center}
\end{figure}

This is a one-sheet hyperboloid, generated by two sets of straight
lines shown in Figure~\ref{f10}. Consider a line segment
\begin{equation}\label{lineseg}
  u\mapsto (\hat r_+,\hat r_-,\hat r_1)+ u(t_+,t_-,t_1)
\end{equation}
through one of the points $\hat p=(\hat r_+,\hat r_-,\hat r_1)\in M_*$
of the hyperboloid. Consider the radius functions $\sqrt{R_i}$,
evaluated as a function of the parameter $u$. If such a function
is affine (has vanishing second derivative) we can set
$(r_1(u),r_2(u))=(\cos\alpha,\sin\alpha)\sqrt{R_i}$ with arbitrary
$\alpha$, to get a straight line in the corresponding hypersurface
in $5$ dimensions. We then call  $(t_+,t_-,t_1)$ an {\it affine direction}
for $R_i$. Along other directions, $R_i$ is strictly concave,
so no decomposition along the segment (\ref{lineseg}) is possible.
For both radius functions, the set of affine directions is a
two-dimensional plane, and thus best described by its normal vector. That
is, $\vec t=(t_+,t_-,t_1)$ is an affine direction for $R_i$ if
$\vec t\cdot\vec A_i=0$, where
\begin{equation}\label{affdir}
  \vec A_1 = \left(\begin{array}{c}
    2 - 3 \hat r_1 - 12 \hat r_-\\
    -2 - 3 \hat r_1 + 12 \hat r_+\\
    -1 + 3 \hat r_- + 3 \hat r_+
  \end{array}\right) \quad
  \vec A_2 = \left(\begin{array}{c}
    -\hat r_1\\
    -\hat r_1\\
    -1 +\hat r_- +\hat r_+
  \end{array}\right).
\end{equation}

Assuming that a convex decomposition along (\ref{lineseg}) is
possible, we thus arrive at a threefold case distinction:
\begin{itemize}
\item The line segment lies entirely in $M_1$.\newline
Then it must be tangent to the hyperboloid $M_*$, and also an
affine direction for $R_1$. The vector $\vec t$ is uniquely determined
up to a factor by these conditions. However, that does not mean that the corresponding
line segment lies in $M_1$, and, in fact, one can show that it
{\it never} does. Hence this case is ruled out.
\item The line segment lies entirely in $M_2$.\newline
This is ruled out analogously.
\item The line segment crosses from $M_1$ into $M_2$.\newline
Then $\vec t$ must be affine for both radius functions. Again,
this determines $\vec t$ to within a factor. But for a proper
decomposition we must have also that the slopes of $\sqrt{R_1}$ and
$\sqrt{R_2}$ match at $u=0$. One can show that this never
happens inside the tetrahedron we discuss, so this case is also
ruled out.
\end{itemize}

We conclude that no point on $M_*$ allows a convex decomposition
inside $\ppt_1$, and the theorem is proved.

\section*{Acknowledgements}

We would like to thank M. Horodecki for discussions.
Funding by the European Union project EQUIP (contract
IST-1999-11053) and financial support from the DFG (Bonn) is
gratefully acknowledged.

\def\pps{PPS}
 \appendix
\section*{Analysis of $\widetilde{\pr}$}\label{sec:app}
In this appendix we give a characterization of the separability
classes ($\widetilde{\tsp},$ $\widetilde{\bsp}_1,$ and
$\widetilde{\ppt}_1$) of $\widetilde{\pr}$ showing that they can
be deduced from those of $\pr$ without any computation.

The intimate relation between the two twirls emerged already in
Lemma~\ref{lem:prprt} where we stated the existence of an
isomorphism between the two algebras spanning the eigenspaces of
$\pr$ and $\widetilde{\pr}$. This isomorphism establishes an
affine mapping $\iota$ between the two eigenspaces that we used to
compute $\ppt_1.$ Due to the inclusion
$\tsp\subsetneq\bsp_1\subsetneq\ppt_1$ it is clear that the same
mapping transports the sets $\tsp$ and $\bsp_1$ to their
counterparts $\widetilde{\tsp}$ and $\widetilde{\bsp}_1.$ The
mapping $\iota$ can be computed by fixing the ordering $\{\idty,
X, V, VXV, XV, VX \}$ for the second algebra and concatenating the
transformations \ref{V2R} and \ref{XV2S} getting
$$\vec{s}=\iota\cdot{\vec{r}}$$ with $$\iota= \left(
\begin{array}{cccccc}
\frac{d-1}{d+1} & 0 & \frac{d+2}{2d+2} & \frac{d+2}{2d+2} & 0 & 0\\
0 & \frac{d+1}{d-1} & \frac{d-2}{2d-2} & \frac{2-d}{2d-2} & 0 & 0\\
\frac{2}{d+1} & -\frac{2}{d-1} & \frac{1}{d^2-1} & -\frac{d}{d^2-1}& 0 & 0\\
\frac{2}{d+1} & \frac{2}{d-1} & -\frac{d}{d^2-1} & \frac{1}{d^2-1} & 0 & 0\\
0 & 0 & 0 & 0 & \frac{\sqrt{3}}{\sqrt{d^2-1}} & 0\\
0 & 0 & 0 & 0 & 0 & \frac{\sqrt{3}}{\sqrt{d^2-1}}
\end{array}\right).
$$

With this mapping we can compute directly the
$\widetilde{\pr}$-projection of the states A to G:

\begin{eqnarray*}
 & A: & \ket{123} \longrightarrow (1/2,1/2,0,0,0) \\
 & B: & \ket{111} \longrightarrow \left( \frac{d-1}{d+1},0,\frac{2}{d+1},0,0\right) \\
 & C: & (\ket{111}-\sqrt{3}\ket{112}+\sqrt{3}\ket{121}-3\ket{122})/4 \\
 & & \qquad \qquad \qquad \longrightarrow \left( \frac{4+5d}{8+8d},\frac{3d-6}{8d-8},\frac{d+2}{4-4d^2},0,0\right) \\
 & D: & \ket{122} \longrightarrow (1,0,0,0,0) \\
 & E: & (\ket{112}-\ket{121})/\sqrt{2} \longrightarrow \left( 0,\frac{d-2}{d-1},\frac{1}{1-d},0,0\right) \\
 & F: & (\ket{123}-\ket{132})/\sqrt{2} \longrightarrow (0,1,0,0,0) \\
 & G: & (\ket{112}-\ket{121}-\sqrt{3}\ket{122})/\sqrt{5}\\
 & & \qquad \qquad \qquad \longrightarrow \left( \frac{3}{5},\frac{2d-4}{5d-5},\frac{2}{5-5d},0,0\right)
\end{eqnarray*}

Applying the transformation to the extremal points and
inequalities of Theorems \ref{thm:tsp}, \ref{thm:bisep} and
\ref{thm:pptex} yields then a characterization of
$\widetilde{\tsp},$ $\widetilde{\bsp}_1$ and $\widetilde{\ppt}_1.$

We omit here the results of these transformations and give the
picture corresponding to Fig.\ref{f4}:

\begin{figure}
\begin{center}
\epsfxsize=8.5cm \epsffile{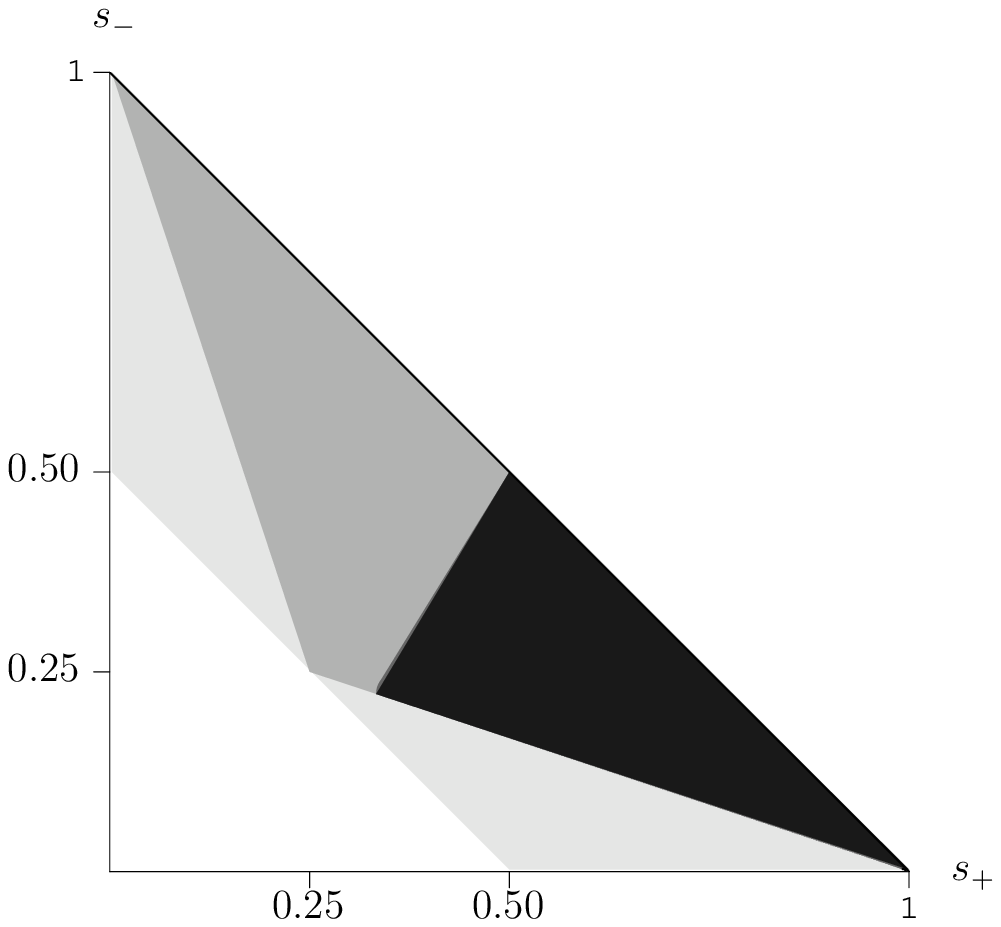} \caption{Sections and
projections of $\widetilde{\tsp}$ and ${\widetilde{\bsp}_1}$
with/onto the $s_+$-$s_-$-plane for $d=3.$ Black: section with
$\widetilde{\tsp},$ dark grey: projection of $\widetilde{\tsp},$
light grey: section with $\widetilde{\bsp}_1.$}\label{f11}
\end{center}
\end{figure}

In contrast to what can be seen in Fig.\ref{f4} the projection of
$\widetilde{\tsp}$ onto the $s_+$-$s_-$-plane differs from its
section with it as one can see in Fig.\ref{f12}:

\begin{figure}
\begin{center}
\epsfxsize=8.5cm \epsffile{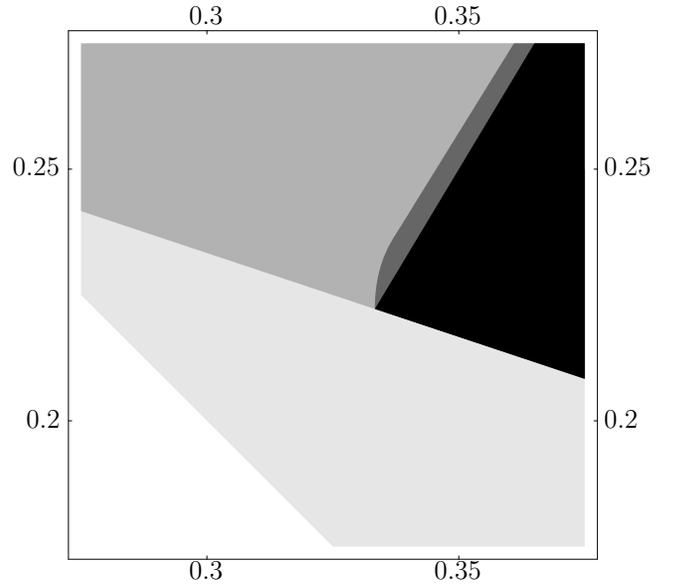} \caption{Zoomed region of
Fig.11.}\label{f12}
\end{center}
\end{figure}


\end{document}